\documentclass[11pt,a4paper]{article}

\usepackage{jheppub}

\usepackage{graphicx, epsfig} 
\usepackage{amsmath,amssymb,amsfonts,dsfont,mathrsfs,amsthm,mathtools}
\usepackage{bm} 
\usepackage{color}
\usepackage[usenames]{xcolor}
\usepackage{hyperref}
\usepackage{physics}
\hypersetup{colorlinks=true,urlcolor=blue,linkcolor=magenta,citecolor=blue,filecolor=blue}
\usepackage[normalem]{ulem}
\usepackage{array}
\usepackage{hyperref}
\usepackage{booktabs}
\usepackage{blindtext}
\usepackage{lipsum}  
\usepackage{slashed}

\allowdisplaybreaks

\usepackage{float}
\usepackage[scr=rsfs]{mathalfa}
\usepackage{bigints}

\newcommand*{\arcsinh}{\operatorname{arcsinh}}

\begin{document}

\title{Partially Massless Condensates and Reentrant NUT/Bolt Phase Transitions in Conformal Gravity}

\author[a]{Crist\'obal Corral,}
\author[b]{Felipe Diaz,}
\author[c]{Borja Diez,}
\author[d]{Eleftherios Papantonopoulos,}
\author[e,f]{Per Sundell}

\affiliation[a]{Departamento de Ciencias, Facultad de Artes Liberales, Universidad Adolfo Ib\'{a}\~{n}ez, Avda. Padre Hurtado 750, 2562340, Vi\~{n}a del Mar, Chile
\vspace{0.1cm}}

\affiliation[b]{Facultad de Ingenier\'ia, Universidad San Sebasti\'an, Santiago 8420524, Chile
\vspace{0.1cm}}

\affiliation[c]{Departamento de F\'{\i}sica, Cinvestav, Av.~IPN 2508, 07360, CDMX, M\'exico}

\affiliation[d]{Physics Division, School of Applied Mathematical and Physical Sciences, National Technical University of Athens, 15780 Zografou Campus, Athens, Greece}

\affiliation[e]{Instituto de Ciencias Exactas y Naturales, Universidad Arturo Prat, \\ Playa Brava 3256, 1111346, Iquique, Chile \& Facultad de Ciencias, Universidad Arturo Prat,\\ Avenida Arturo Prat Chacón 2120, 1110939 Iquique, Chile
\vspace{0.1cm}}

\affiliation[f]{Facultad de Ciencias, Universidad Arturo Prat, Avenida Arturo Prat Chac\'on 2120, 1110939, Iquique, Chile
\vspace{0.1cm}}

\vspace{0.3cm}
\emailAdd{cristobal.corral@uai.cl}
\emailAdd{felipe.diaz@uss.cl}
\emailAdd{borja.diez@cinvestav.mx}
\emailAdd{lpapa@central.ntua.gr}
\emailAdd{psundell@unap.cl}

\abstract{
We study non-Einstein Bach-flat Taub-NUT/Bolt instantons with weakened AdS boundary conditions. In particular, the next-to-leading-order coefficient in the Fefferman--Graham expansion is nonvanishing, and it sources the partially massless response in the dual conformal field theory. As in the Einstein case, the holographic stress tensor of these instantons represents a perfect fluid with vorticity. Remarkably, we identify a non-Einstein Taub-Bolt branch supporting a nonvanishing partially massless response at zero source, while preserving the full symmetry of the squashed conformal boundary. The additional holographic degrees of freedom enrich the thermodynamic phase structure, allowing the conventional NUT/Bolt transition to disappear or give way to reentrant transitions. We further show that non-Einstein Bolts can possess negative thermodynamic volume and establish the absence of negative modes of a conformally coupled probe scalar on regular Bach-flat NUT backgrounds.

}


\maketitle

\section{Introduction}

Conformal gravity is an interesting theory that, in addition to diffeomorphism invariance, remains invariant under Weyl rescalings of the metric. Its action is given by the only conformal invariant in four dimensions constructed solely from the metric and its derivatives: the square of the Weyl tensor. It has been considered an ultraviolet completion of Einstein gravity, since its higher-derivative nature improves the graviton propagator in the ultraviolet regime, even though quantum corrections break conformal symmetry~\cite{Adler:1982ri}. In fact, it has been shown that conformal gravity is renormalizable at one loop~\cite{Stelle:1976gc,Stelle:1977ry,Buchbinder:1984ebp,Buchbinder:2026bfn}, however, it is not unitary. Nevertheless, it provides a suitable ground for studying holography from a top-down viewpoint, since it appears naturally in twistor-string theory~\cite{Witten:2003nn,Berkovits:2004hg}, whose low-energy limit possesses conformal gravity in its bosonic sector~\cite{Berkovits:2004jj}. Additionally, in four dimensions, conformal gravity reduces to Einstein-AdS gravity at the level of classical solution space and on-shell action by imposing Neumann boundary conditions on the Fefferman--Graham expansion \cite{Maldacena:2011mk,Anastasiou:2016jix},\footnote{For higher-dimensional extensions, see \cite{Anastasiou:2020mik,Anastasiou:2026jrt}.} which eliminate the ghost modes, as it was suggested in Refs.~\cite{Lu:2011ks,Hell:2023rbf}. Thus, conformal gravity is an interesting theoretical laboratory for extending the original correspondence between anti-de Sitter (AdS) gravity and conformal field theories (CFT) \cite{Maldacena:1997re,Witten:1998qj,Gubser:1998bc} to models with extra holographic sources, well-defined Einstein limits, and possibly other values of the cosmological constant, though our focus will be on the AdS case. The model is also attractive from a practical perspective since its classical action is finite and obeys the variational principle for holographic sources in asymptotically locally AdS spacetimes without the need for any boundary term~\cite{Grumiller:2013mxa}.

The classical solution space of conformal gravity, whose field equations are fourth-order in derivatives, contains Einstein branches with dynamically generated cosmological constants.
When linearised around maximally symmetric backgrounds, the linearised spectrum consists of six propagating local degrees of freedom forming i) a single indecomposable non-unitary irrep of the Poincaré algebra containing a unitary graviton on flat backgrounds; ii) a unitary graviton and a non-unitary partially massless graviton on AdS backgrounds; and iii) a unitary graviton and a unitary partially massless graviton on de Sitter (dS) backgrounds.
More specifically, linearised around a Minkowski background, the theory propagates two massless spin-2 fields and one massless spin-1 field, giving a total of six degrees of freedom~\cite {Riegert:1984hf}. On global AdS, however, the particle spectrum of conformal gravity consists of a massless spin-2 field and a partially massless spin-2 field with $\pm2$ and $\pm1$ helicities, giving six degrees of freedom as well~\cite{Lu:2011zk,Lu:2011ks,Deser:2012euu,Deser:2012qg} (see also~\cite{Boulanger:2023lgd,Boulanger:2024qrn}). Independent degree-of-freedom counting using the Hamiltonian formalism confirmed these results~\cite {Chen:2018dqz}. Importantly, the higher-derivative nature of conformal gravity implies that its field equations do not fix the first four coefficients in the Fefferman--Graham expansion. The resulting non-Einstein modes facilitate relaxed AdS boundary conditions at the conformal boundary, providing an interesting setup for holography~\cite{Grumiller:2013mxa}. The issue of unitarity is particularly relevant in the holographic context. In AdS, the additional partially massless spin-two mode is dual to an operator of conformal dimension $\Delta=2$, violating the unitarity bound for spin-two primaries in three-dimensional CFTs. In contrast, partially massless fields can furnish unitary representations of the de Sitter isometry group, although this does not establish the unitarity of the full conformal gravity theory or its putative holographic dual \cite{Dolan:2001ih}. 

In four dimensions, obtaining analytic higher-curvature corrections to rotating black holes is quite challenging. However, the Weyl invariance of conformal gravity renders this higher-curvature theory unique, since it admits analytic, stationary, non-Einstein solutions.\footnote{Recently, a non-Einstein solution to conformal gravity was found in higher dimensions~\cite{Guajardo:2026vzh}.} For instance, the Bach-flat extension of the Kerr--Newman black hole was obtained in Ref.~\cite{Mannheim:1990ya} and further analyzed in~\cite{Asuncion:2025cxw}. Its static limit is the Riegert black hole, the unique spherically symmetric solution of conformal gravity~\cite{Riegert:1984zz}; the latter was shown to be conformally related to the Schwarzschild black hole~\cite{Schmidt:1999vp}. The whole family of Bach-flat Plebanski--Demianski metrics was found in Ref.~\cite{Liu:2012xn}, including the Kerr--Newman-NUT-AdS spacetime, whose conserved charges were studied in Ref.~\cite{Corral:2025npd} using the Noether--Wald formalism~\cite{Lee:1990nz,Wald:1993nt,Iyer:1994ys,Wald:1999wa}. Intriguingly,  in Ref.~\cite{Mannheim:1988dj} it was argued that the modification to the Newtonian potential induced by the linear mode in conformal gravity could explain the observed galactic rotation curves, although this mechanism remains debated~\cite{Edery:1997hu,Yoon:2013rxa,Mannheim:2015gba,Hobson:2021vzg,Mannheim:2021fql,Mannheim:2021osa}. 

Among all the stationary metrics in conformal gravity, instantons deserve particular attention. They represent Euclidean stationary saddle points in the path integral formulation of conformal gravity and allow analytic computation of vacuum persistence between topologically inequivalent vacua~\cite {Strominger:1984zy}. Since conformal gravity is a quadratic form in the Weyl tensor, self-dual gravitational instantons saturate a Bogomol’nyi--Prasad--Sommerfield bound, providing a global extremum labeled by the Chern-Pontryagin index~\cite{Deser:1983wb}. More recently, the Bach-flat extension of the self-dual Taub-NUT and Eguchi-Hanson instantons was studied in Ref.~\cite{Corral:2021xsu}, and their coupling to nonlinear conformal matter was further analyzed in~\cite{Corral:2025npd}. The former has relaxed AdS boundary conditions and it is particularly interesting for holography, since its Einsteinian counterpart gives rise to holographic fluids with nontrivial vorticity~\cite{Caldarelli:2011idw,Leigh:2011au,Leigh:2012jv,Mukhopadhyay:2013gja,Kalamakis:2020aaj}. Moreover, the boundary of the Euclidean Taub-NUT-AdS instanton is a squashed three-sphere whose squashing parameter is given in terms of the NUT charge. The latter is not conformally flat, as its boundary Cotton tensor is nonvanishing for generic values of the NUT parameter. This provides an interesting arena for studying holographic CFTs on compact nonconstant-curvature spaces with broken isometries due to the squashing~\cite{Bobev:2016sap,Bobev:2017asb,Bueno:2018yzo,Bueno:2020odt,Bobev:2023lkx,Parmentier:2026aqh}. Thus, it is natural to ask what is the role of the linear mode of conformal gravity in holographic duals to asymptotically locally AdS gravitational instantons.

The thermodynamics of gravitational instantons with squashed three-sphere boundaries has attracted considerable interest in connection with the AdS/CFT correspondence; see, for instance, \cite{Hawking:1998ct,Dowker:1998pi,Emparan:1999pm,Chamblin:1998pz,Birmingham:2001vd,Clarkson:2002uj,Martelli:2011fw,Martelli:2012sz,Johnson:2014pwa,Johnson:2014xza, Bobev:2016sap,Bueno:2018yzo, Bordo:2019slw,Diez:2026qjt}. In particular, Hartnoll and Kumar \cite{Hartnoll:2005yc} investigated the large-$N$ $O(N)$ vector model on a squashed sphere and compared its thermal behaviour with the semiclassical Einsteinian gravitational partition function constructed from Euclidean Taub-NUT and Taub-Bolt saddles. Even though the critical $O(N)$ vector model has been conjectured to be dual to AdS higher spin gravity at large-$N$~\cite{Sezgin:2002rt,Klebanov:2002ja}, it was shown that Euclidean Einstein-AdS solutions already capture certain qualitative attributes of the putative dual theory on the squashed three-sphere~\cite{Hartnoll:2005yc}. In particular, the vector model's free energy reproduces several characteristic features of the renormalized Einstein-AdS on-shell action, including its maximum at the round sphere, its behaviour at extreme squashings, and its linear dependence on the squashing parameter in the large-squashing regime. Moreover, the Einstein Taub-NUT and Taub-Bolt branches capture the behaviour of the vector-model free energy in different regions of parameter space. However, an important discrepancy remains. While Einstein gravity exhibits a first-order Hawking-Page-type transition between NUT and Bolt saddles, the vector model interpolates smoothly between the corresponding regimes without an analogous transition.

This discrepancy suggests that the gravitational theory's thermodynamic behaviour may depend on degrees of freedom beyond those of Einstein gravity, motivating investigation of higher-derivative extensions. In this regard, conformal gravity provides a particularly interesting setup, since the Bach-flat extension of Taub-NUT/Bolt spaces has been found analytically~\cite{Corral:2021xsu,Corral:2025npd}, unlike in other four-dimensional higher-curvature gravity theories~\cite{Bueno:2018uoy}. This allows us to investigate their thermodynamic properties.

In this work, we study the thermodynamics and holographic responses of the Bach-flat Taub-NUT instantons with relaxed AdS boundary conditions. To this end, we explicitly compute their coefficients in the Fefferman--Graham expansion and show that they take a perfect-fluid form. The boundary metric is not deformed by the non-Einstein mode of conformal gravity. However, its holographic response gets nontrivial contributions from it. The mass and angular momentum of the solution are obtained from the holographic data, and differ from the Noether--Wald potential used in  Refs.~\cite{Corral:2021xsu,Corral:2025npd} by a $\dd$-exact term proportional to the first Chern number of the Hopf bundle, with a coefficient set by the partially massless source. The holographic stress tensor satisfies a modified Ward identity as a consequence of the partially massless response. Additionally, we verify that it represents a perfect fluid with nontrivial vorticity as in the Einstein case.

Interestingly, there exists a curve in the parameter space where the partially massless response condenses. This curve is analyzed in detail, and it is found that the condensate does not break any symmetry whatsoever. This behaviour resembles the Nambu-Jona-Lasinio model \cite{Nambu:1961tp} in which the quark condensate preserves global isospin and baryon-number symmetry. The thermodynamic phase structure is analyzed in detail. It is found that conformal gravity allows for reentrant phase transitions between the Bach-flat Taub-NUT and Taub-Bolt spaces, which is absent in their Einsteinian counterpart. Then, we study scalar perturbations by analyzing the eigenvalue problem of a probe conformal scalar field on the Bach-flat Taub-NUT geometry. We find that there are no negative eigenvalue modes of the conformal scalar fields, extending the results of Ref.~\cite{Hartnoll:2005yc} consistently with the symmetric phase of the $O(N)$ model. Finally, we compute the thermodynamic volume dual to the AdS radius, which appears as an integration constant.

The manuscript is organized as follows. In Sec.~\ref{sec:thesolution}, we review the Bach-flat Taub-NUT and Taub-Bolt solutions of conformal gravity. Section~\ref{sec:holography} determines their asymptotic data and holographic one-point functions and investigates the emergence of partially massless condensates. In Sec.~\ref{sec:thermo}, we examine the thermodynamic phase structure, including reentrant transitions, the stability of conformally coupled scalar perturbations, and the thermodynamic volume. Finally, Sec.~\ref{sec:discussion} summarizes our results and discusses possible future directions.

\section{Conformal Gravity and Taub-NUT/Bolt Geometries\label{sec:thesolution}}
Our starting point is conformal gravity in the Euclidean signature described by the action 
\begin{align}\label{action}
    I[g] = -\alpha\int_\mathcal{M} \dd^4x\sqrt{|g|}\;W_{\mu\nu\rho\sigma}W^{\mu\nu\rho\sigma}~,
\end{align}
where $W_{\mu\nu\rho\sigma}$ is the Weyl tensor and $\alpha$ is dimensionless.
Moreover, the action has a proper holographic description and is finite without any boundary term~\cite{Grumiller:2013mxa}. This means that it has a well-posed variational principle with Dirichlet boundary conditions for the holographic sources at the conformal boundary.
The action is invariant under diffeomorphisms and local Weyl transformations, the latter acting on the metric as $g_{\mu\nu}\to e^{2\sigma(x)}g_{\mu\nu}$, under which the Weyl tensor transforms as $W_{\mu\nu\rho\sigma}\to e^{2\sigma(x)}W_{\mu\nu\rho\sigma}$. The Weyl tensor contains important information: its electric part governs tidal forces due to gravity, while its magnetic part governs tidal frame dragging. These quantities define conserved charges on asymptotically locally AdS spacetimes via the Ashtekar-Magnon-Das method~\cite{Ashtekar:1984zz,Ashtekar:1999jx}. In this prescription, suitable rescaling of tensorial quantities makes the conserved charge formulas expressible in terms of the Weyl tensor. Additionally, the stability properties of Weyl geometry have also been investigated in Ref.~\cite{Kouniatalis:2025pxs}.

The variation of the action~\eqref{action} with respect to the metric yields the fourth-order field equations
\begin{align}\label{Bach-eom}
    B_{\mu\nu} = 0~,\qquad \text{where}\qquad B_{\mu\nu}:= \left(\nabla^\rho\nabla^\sigma +\frac12 R^{\rho\sigma}\right)W_{\mu\rho\nu\sigma}~,
\end{align}
is the Bach tensor, which is symmetric, traceless, and covariantly conserved. It is worth mentioning that all Einstein spaces are Bach-flat. Therefore, they automatically satisfy the conformal gravity equations of motion. However, the converse is not necessarily true, and conformal gravity admits non-Einstein solutions with a richer asymptotic structure. Unlike Einstein gravity, the action contains no explicit cosmological constant term, and solutions with different effective curvature scales are allowed.

To make the Einstein and non-Einstein contributions explicit, we introduce an arbitrary reference length $\ell$ and define $E_{\mu\nu}:=R_{\mu\nu}+3\ell^{-2}g_{\mu\nu}$. Although this scale does not appear in the original conformal-gravity action, the resulting decomposition is an exact off-shell identity, with all explicit dependence on $\ell$ canceling in the sum. For the solutions considered here, we choose $\ell>0$ to coincide with their asymptotic AdS radius, so that $E_{\mu\nu}$ measures the departure from the corresponding Einstein geometry,
\begin{align}
    I = -\frac{4\alpha}{\ell^2}\int_{\cal M}\dd^4x\sqrt{g}\left[R+\frac{6}{\ell^2} + \frac{\ell^2}{4}\left({\cal E}_4 + 2E_{\mu\nu}E^{\mu\nu} - \frac23E^2\right)\right]~,
\end{align}
where
\begin{align}
    {\cal E}_4 = R_{\mu\nu\rho\sigma}R^{\mu\nu\rho\sigma}-4R_{\mu\nu}R^{\mu\nu} + R^2~,
\end{align}
is the four-dimensional Euler density. Indeed, for Einstein metrics $E_{\mu\nu} = 0$ and the action reduces to the renormalized action of~\cite{Olea:2005gb} upon identifying \begin{align}
\alpha = \frac{\ell^2}{64\pi G}~,
\end{align}
which matches the standard renormalized action of Ref.~\cite{deHaro:2000vlm} modulo a finite term proportional to the Euler characteristic of the manifold~\cite{Anastasiou:2020zwc}. 
The equivalence of this action with the one obtained through standard holographic renormalization~\cite{deHaro:2000vlm} was established in Ref.~\cite{Anastasiou:2020zwc}, ensuring a well-posed variational principle for the holographic sources and a finite on-shell action under the appropriate boundary conditions. 
Moreover, Ref.~\cite{Anastasiou:2016jix} exploited this connection to extend Maldacena's argument~\cite{Maldacena:2011mk} on the equivalence between Einstein gravity and conformal gravity in the Einstein sector.

We are particularly interested in the Euclidean Taub-NUT/Bolt instantons, including their non-Einstein Bach-flat branches. These spaces can be described in terms of Hopf fibrations over $\mathbb{S}^2$, with line elements of the form~\cite{Taub:1950ez,Newman:1963yy,Page:1985bq}
\begin{align}
    \dd s^2 = f(r)(\dd\tau + 2n\cos\theta \dd\phi)^2 + \frac{\dd r^2}{f(r)} + (r^2-n^2)\left(\dd\theta^2+\sin^2\theta\dd\phi^2\right)~,
\end{align}
with $\tau \sim \tau + \beta$.
Before going into the details of this usage in the context of conformal gravity, a few general remarks of relevance are the following. 
These families of spaces are cohomogeneity-one metrics whose principal orbits (the constant-$r$ slices) are squashed three-spheres written as a Hopf fibration, with $\tau$ being the fiber of the $U(1)$ bundle and $n$ setting the twist. The one-form $\cos\theta \dd\phi$ is not globally defined on the south-pole and north-pole axes, where the Misner strings are located~\cite{Misner:1963fr}. More generally, the position of the string can be changed by shifting $\tau\to\tau + 2nC\phi$ with $C=\{0,1,-1\}$ such that the aforementioned one-form becomes $2n(\cos\theta + C)\dd\phi$. This shift has no effect in Einstein gravity, and as we shall show, the same holds in conformal gravity. The strings can be removed by working in two patches, $\tau_\pm = \tau \pm 2n\phi$, regular around $\theta = 0$ and $\theta=\pi$, respectively. The transition function $\tau_+  - \tau_- = 4n\phi$ is consistent with the periodicity of $\phi$ only if $\tau$ has period $8\pi n$. In that case, $\dd\tau + 2n\cos\theta\dd\phi$ is, up to a $U(1)$ gauge transformation, $2n$ times the global connection one-form of the Hopf bundle, and the constant-$r$ slices are smooth squashed three-spheres. 

The $U(1)$ action is generated by the Euclidean Killing vector $\partial_\tau$, and its norm is $g(\partial_\tau,\partial_\tau)=f(r)$~. The Euclidean geometry therefore closes off where $f$ vanishes, since the orbits of $\partial_\tau$ shrink to zero size there. The closing is smooth without a conical singularity if there is some $r_0$ being a simple zero of $f$ and the period of $\tau$ is $\beta=4\pi/f'(r_0)$. In summary, the regularity conditions are
\begin{align}\label{regconditions}
    \tau \sim \tau+\beta~,\qquad \beta = \frac{4\pi}{f'(r_0)}~,\qquad f(r_0)=0~.
\end{align}

The Taub-NUT and Taub-Bolt geometries arise from the two ways in which the regularity conditions~\eqref{regconditions} can be obeyed, which are distinguished by the topology of the set of fixed points of the Killing vector $\partial_\tau$, as follows: 
\begin{itemize}
    \item NUT:
If $r_0 = n$, the transverse two-sphere of radius $\sqrt{r^2-n^2}$ shrinks together with the fiber, and the whole $\mathbb{S}^3$ collapses to a point. This space, with a zero-dimensional set of fixed points, is referred to as a \textit{nut} and is locally $\mathbb{R}^4$. The absence of conical singularities requires $f'(n) = 1/2n$ so that $\beta=8\pi n$, and there is no Misner string in this space. 
\item Bolt: If  $r_0 = r_b > n$, only the fiber shrinks, while the two-sphere keeps a finite radius $\sqrt{r_b^2-n^2}$. Such a space, with the set of fixed points being a two-sphere, is called a \textit{bolt}, and near the sphere the space is locally $\mathbb{R}^2\times \mathbb{S}^2$. In this case, the smoothness condition $\beta=4\pi/f'(r_b)$ is independent of the Misner condition $\beta = 8\pi n$, so imposing both simultaneously fixes the radius $r_b$ in terms of the remaining parameters governed by dynamics. The resulting space is an $\mathbb{R}^2$ bundle over $\mathbb{S}^2$ which is topologically $\mathbb{CP}^2$ with a point being removed.\footnote{More generally, one can take the period of $\tau$ to be $8\pi n/k$ with $k$ a positive integer. The constant-$r$ slices are then lens spaces $S^3/\mathbb{Z}_k$, i.e. circle bundles over $S^2$ with first Chern number $k$. For the NUT, smoothness at $r=n$ requires $k=1$ otherwise the NUT becomes a $\mathbb{R}^4/\mathbb{Z}_k$ orbifold point, whereas for the bolt, any $k$ yields a smooth space, an $\mathbb{R}^2$ bundle over the bolt two-sphere whose circle wind $k$ times around it (the complex line bundle of degree $-k$ over $\mathbb{CP}^1$). For instance, $k=2$ gives $T^\ast S^2$, the topology of the Eguchi--Hanson instanton. In this work we restrict to the $k=1$ case. }
\end{itemize}
Within this class of geometries, Refs.~\cite{Liu:2012xn,Corral:2021xsu} found NUTs and bolt spaces that are non-Einstein and Bach-flat and solve the conformal gravity field equations. The metric function for both families is
\begin{align}\label{fsol}
    f(r) = \frac{1}{r^2-n^2}\Big[r^2 + n^2 -2mr + \tfrac{1}{\ell^2}\left(r^4-6n^2r^2 - 3n^4\right)+ a\left(r^2+\tfrac13 n^2\right)+br^3\Big]~,
\end{align}
which introduces an additional set of integration constants $(a, b,m, \ell) \in \mathbb{R}^4$ subject to the Bach-flat condition
\begin{align}\label{BachFlatConstraint}
    6bm+ a\left(a+2-\frac{8n^2}{\ell^2}\right) =0~.
\end{align}
In particular $\ell$, controls the asymptotic value of the curvature, playing the role of the cosmological constant. Thus, in contrast to Einstein gravity, conformal gravity does not contain a cosmological constant in its field equations. Instead, an effective cosmological constant arises as an integration constant and can have either sign, allowing for asymptotically AdS, dS, or flat solutions. In this work, we focus on the asymptotically AdS branch, for which $\ell^2>0$.

The metric is Einstein only for $a=b=0$, as can be seen from the Ricci scalar
\begin{align}
    R = -\frac{12}{\ell^2}-\frac{2a+6br}{r^2-n^2}~,
\end{align}
and the norm of the traceless Ricci tensor\footnote{As we are in Euclidean signature, the vanishing of the norm of a tensor implies the vanishing of the tensor itself.}
\begin{align}
    \left(R_{\mu\nu}-\frac14 g_{\mu\nu}R\right)\left(R^{\mu\nu}-\frac14 g^{\mu\nu}R\right)= \left[\frac{br\left(r^2+3n^2\right) + a\left(r^2+\frac{n^2}{3}\right)}{(r^2-n^2)^2}\right]^2~.
\end{align}

The parameter $b$ controls a term growing linearly in $r$ in the infrared, $f \simeq \frac{r^2}{\ell^2} + br + {\mathcal O}(1)$, and the Einstein AdS Taub-NUT/Bolt family is recovered smoothly by setting $b=0$ and $a=0$. The regularity analysis above applies unchanged, since it only depends on the behaviour of $f$ near its zero. Demanding that the geometry closes off at a NUT, i.e. $f(n) = 0$ and $f'(n)=1/(2n)$, imposes the following constraints on the integration constants
\begin{align}\label{NUTregularityconds}
    a=-3bn=:a_{\rm NUT}~,\qquad m = n-\frac32 bn^2 - \frac{4n^3}{\ell^2}=:m_{\rm NUT}~.
\end{align}
Substituting these relations into the metric function in \eqref{fsol} yields
\begin{align}\label{f-nut}
    f_{\rm NUT}(r) = \frac{r-n}{r+n}\left[  1+b(r-n)+\frac{1}{\ell^2}(r-n)(r+3n)\right]~.  
\end{align}Furthermore, the absence of conical singularities at the NUT requires the Euclidean time coordinate to be periodically identified with period
\begin{equation}
	\beta_{\rm NUT}= \frac{4\pi}{f_{\rm NUT}'(n)}=8\pi n~.
\end{equation}
The resulting geometry is globally self-dual or anti-self-dual, depending on the orientation, meaning that the Weyl tensor satisfies $W_{\mu\nu\lambda\rho}=\pm\tilde{W}_{\mu\nu\lambda\rho}$, where $\tilde{W}_{\mu\nu\lambda\rho}=\tfrac{1}{2}\varepsilon_{\mu\nu\alpha\beta}W^{\alpha\beta}{}_{\lambda\rho}$ is the dual Weyl tensor with $\varepsilon_{\mu\nu\alpha\beta}$ being the Levi-Civita tensor. Then, the action can be evaluated directly in terms of the Chern-Pontryagin index since, for (anti-) self-dual configurations, one finds 
\begin{align}
I = \mp \alpha\int_{\mathcal{M}}\dd^4x\sqrt{|g|}\;\tilde{W}_{\mu\nu\lambda\rho}W^{\mu\nu\lambda\rho} =  \mp16\pi^2\alpha P_1[\mathcal{M}]\,.
\end{align}
Here, the off-shell identity $\tilde{W}^{\mu\nu}{}_{\lambda\rho}W^{\lambda\rho}{}_{\mu\nu}=\tilde{R}^{\mu\nu}{}_{\lambda\rho}R^{\lambda\rho}{}_{\mu\nu}$ has been used and the Chern-Pontryagin index is defined as
\begin{align*}
 P_1[\mathcal{M}] = \frac{1}{16\pi^2}\int_{\mathcal{M}}\dd^4x\sqrt{|g|}\;\tilde{R}^{\mu\nu}{}_{\lambda\rho}R^{\lambda\rho}{}_{\mu\nu}\,.
\end{align*}
In fact, Refs.~\cite{Deser:1983wb,Strominger:1984zy} show that self-dual instantons in conformal gravity saturate a BPS bound, whose value is determined by the Chern-Pontryagin index above.\footnote{In the presence of boundaries, as in the case of Taub-NUT, the Chern-Pontryagin index is not a topological invariant. Rather, one needs to include the Atiyah-Patodi-Singer $\eta$-invariant that accounts for the spectral asymmetry~\cite{Atiyah:1975jf,Atiyah:1976jg,Atiyah:1976qjr}.} In particular, in the case of the self-dual Bach-flat Taub-NUT metric, it yields~\cite{Corral:2021xsu}
\begin{align}\label{Inut}
    I_{\rm NUT} = -32\pi^2\alpha \left[\left(1-\frac{4n^2}{\ell^2}\right)^2+2b^2n^2\right]~.
\end{align}

On the other hand, for the Taub-Bolt solution with weakened AdS boundary conditions, the fixed-point set is located at $r=r_b>n$, where $f(r_b)=0$. This fixes the integration constant $m$ as
\begin{align}\label{mbolt}
    m = \frac{1}{2r_b}\left[r_b^2+n^2+\frac{1}{\ell^2}\left(r_b^4-6n^2r_b^2-3n^4\right)+a\left(r_b^2+\frac13 n^2\right)+br_b^3\right] =: m_{\rm bolt}~.
\end{align}
The absence of a conical singularity at the bolt fixes the period of the Euclidean time coordinate, identified with the inverse temperature, as
\begin{equation}\label{boltbeta}
	\beta_{\rm gen bolt}=\frac{12\pi r_b(r_b^2-n^2)}{6br_b^3+a(3r_b^2-n^2)+3(r_b^2-n^2)+\frac{9}{\ell^2}(r_b^2-n^2)^2}~.
\end{equation}
In this case, the corresponding on-shell Euclidean action is given by
\begin{equation}\label{Igenbolt}
	\begin{split}
		I_{\rm gen bolt}={}&-\frac{16\pi \alpha\beta_{\rm genbolt}}{9r_b\ell^4(r_b^2-n^2)^2}\left\{9r_b^8+18b\ell^2r_b^7+9(b^2\ell^4+\ell^2(a+2)-2n^2)r_b^6\right.\\
        &+9b\ell^2[\ell^2(a+2)-4n^2]r_b^5+[3\ell^4(3b^2n^2+a^2+3a+3)-3(7a+24)\ell^2n^2+36n^4]r_b^4\\
        &-6b\ell^2n^2[(3-a)\ell^2-3n^2]r_b^3+[(-a^2-12a-18)\ell^4+15n^2(a+6)\ell^2-54n^4]n^2r_b^2\\
        &-3ab\ell^4n^4r_b\left.-3n^4(\ell^2-n^2)[(-3-a)\ell^2+9n^2]\right\}
	\end{split}
\end{equation}
where we recall that the integration constants are subject to the constraint in Eq.~\eqref{BachFlatConstraint}, together with the bolt condition~\eqref{mbolt}.
Furthermore, removing the Misner string imposes the additional identification $\beta_{\rm bolt}=8\pi n$. Combined with the condition of regularity at the bolt, this leads to the relation
\begin{align}\label{boltregularityconds}
    a = \frac{3}{3r_b^2-{n^2}}\left[\left(r_b^2-n^2\right)\left(\frac{r_b}{2n}-1\right)-\frac{3}{\ell^2}\left(r_b^2-n^2\right)^2-2br_b^3\right] =: a_{\rm bolt}~.
\end{align}
In what follows, we distinguish between Taub-Bolt instantons satisfying only the local regularity condition at the fixed-point set and those that additionally satisfy the global regularity condition associated with the removal of Misner strings. For the latter configurations, the on-shell action takes the form
\begin{align}\label{Iboltsmooth}
    I_{\rm bolt}={}&-128\alpha\pi^2n\left\{ \frac{a_{\rm bolt}r_b(3r_b^2-n^2)}{3}\left[\frac{a_{\rm bolt}}{3(n^2-r_b^2)^2}+\frac{bn^2r_b+br_b^3-n^2+r_b^2}{r_b^2(n^2-r_b^2)^2}+\frac{1}{r_b^2\ell^2}\right]\right. \nonumber\\&\left.+\frac{n^4+r_b^2n^2(b^2r_b^2-2br_b-2)+r_b^4(br_b+1)^2}{r_b(n^2-r_b^2)^2}+\frac{2(br_b^3-2n^2+r_b^2)}{r_b\ell^2} +\frac{(r_b^4+3n^4)}{r_b\ell^4}\right\}~,
\end{align}
which depends on four free parameters, viz $(\ell,n,r_b,b)$ modulo the conditions from Bach-flatness and global smoothness. 

The Euclidean Bach-flat Taub-NUT and Taub-Bolt instantons are globally and asymptotically self-dual, respectively. They provide distinct smooth fillings of a squashed three-sphere at conformal infinity. Although they share the same asymptotic structure, they differ in the topology and dimensionality of their set of fixed points. The integration constants provided by the additional modes of conformal gravity allow for weakened AdS boundary conditions beyond the standard Einstein branch. These features will play an important role in the holographic description of the solutions, where the boundary geometry and the additional asymptotic data determine the corresponding conserved quantities and thermodynamic properties.

\section{Holographic Aspects\label{sec:holography}}

In four-dimensional conformal gravity, the asymptotic expansion contains additional independent boundary data compared to Einstein gravity, reflecting the presence of higher-derivative degrees of freedom. In particular, the first subleading coefficient of the metric is not required to vanish and acts as an additional source in the dual field theory. To investigate the holographic properties of the Euclidean Taub-NUT and Taub-Bolt instantons with weakened AdS boundary conditions, we first determine their asymptotic structure in the Fefferman–Graham (FG) gauge.

\subsection{Asymptotic data and holographic one-point functions} 

The standard prescription to obtain the holographic data in asymptotically locally AdS spacetimes starts by writing the metric in the FG form~\cite{FeffermanGraham1985,FeffermanGraham2012}
\begin{align}\label{FG-gauge}
    \dd s^2 = \frac{\ell^2}{z^2}\left[\dd z^2+\gamma_{ij}(z,x^i)\dd x^i\dd x^j\right]~,
\end{align}
where the conformal boundary is located at $z=0$, and the induced metric admits the expansion
\begin{align}\label{FG2}
    \gamma_{ij}(z,x^i) = \gamma_{ij}^{(0)} + \frac{z}{\ell}\gamma_{ij}^{(1)}+\frac{z^2}{\ell^2}\gamma_{ij}^{(2)}+\dots~.
\end{align}
Then, one expands the field equations asymptotically and solves them order by order in the holographic coordinate to determine the coefficients in the FG expansion. A further analysis of the action and variations thereof allows one to obtain the divergences of the action, which can be removed by adding counterterms~\cite{Balasubramanian:1999re,Emparan:1999pm,deHaro:2000vlm,Bianchi:2001kw,Skenderis:2002wp}. Then, from the variational principle at the conformal boundary, one can read the sources and holographic one-point functions directly.  

The holographic dictionary of four-dimensional conformal gravity was established in Ref.~\cite{Grumiller:2013mxa}. To do so, relaxed AdS boundary conditions were introduced, namely, weaker than the standard Starobinsky conditions on the FG expansion in Einstein gravity.\footnote{The absence of the linear term in the FG expansion of four-dimensional Einstein-(A)dS gravity goes back to Starobinsky’s asymptotic analysis of de Sitter space~\cite{Starobinsky:1982mr}, and was subsequently incorporated into the general FG expansion of conformally compact Einstein metrics~\cite{FeffermanGraham1985,FeffermanGraham2012}.} In conformal gravity, however, the next-to-leading order coefficient $\gamma^{(1)}_{ij}$ in the FG expansion is not constrained to vanish by virtue of the Bach-flat equations~\eqref{Bach-eom}. In fact, the latter provides an independent holographic source, with a well-posed variational principle, i.e.
\begin{align}
    \delta I = \int_{\partial\cal M}\dd^3x\sqrt{\gamma^{(0)}}\left(\tau^{ij}\delta \gamma^{(0)}_{ij} +  P^{ij}\delta\gamma^{(1)}_{ij}\right)~.
\end{align}
Here, $\tau^{ij}$ denotes the conformal-gravity analog of the Brown-York response and $P^{ij}$ is the partially massless response, both of which are thus finite.

As shown in Ref.~\cite{Grumiller:2013mxa}, the holographic responses are best organized in terms of the electric and magnetic parts of the Weyl tensor, defined as\footnote{Strictly speaking, the electric and magnetic parts of the Weyl tensor are conventionally defined with respect to a timelike unit vector in Lorentzian spacetime. In the present Euclidean setting, the decomposition is instead performed with respect to the unit normal to the constant-radial-coordinate hypersurfaces. We retain the standard terminology of ``electric'' and ``magnetic'' Weyl tensors, as is customary in holographic analyses.} 
\begin{align}
\mathcal E_{\mu\nu}=W_{\mu\alpha\nu\beta}N^\alpha N^\beta~, \qquad \mathcal B_{\mu\nu\rho}= W_{\alpha\mu\nu\rho}N^\alpha ~,
\end{align}
respectively, where $N^\mu\partial_\mu = -\frac{z}{\ell}\partial_z$ is the outward-pointing unit normal to a codimension-1 cutoff surface $z={\rm const}$; it can also be defined in terms of the FG coefficients~\cite{Anastasiou:2020mik}. Their asymptotic expansions are denoted by
\begin{align}
    {\cal{E}}_{ij} ={} {\cal{E}}_{ij}^{(2)} + \frac{z}{\ell}{\cal E}^{(3)}_{ij} + \dots~, \qquad {\cal B}_{ijk} = \frac{\ell}{z}{\cal B}^{(1)}_{ijk} + {\cal B}^{(2)}_{ijk} + \dots~,
\end{align}
where their form in terms of the FG coefficients was derived in Ref.~\cite{Grumiller:2013mxa}. For the sake of completeness, we include them here; they are 
\begin{equation}
    \label{eq:FGcoeffWeyl}
    \begin{aligned}
        \mathcal{B}^{(1)}_{ijk}
        ={}& \frac{1}{2\ell}\left( D_j \psi^{(1)}_{ik} - \frac{1}{2}\gamma^{(0)}_{ij} D^{l}\psi^{(1)}_{kl} \right) - j \leftrightarrow k\,, \\
        \mathcal{E}^{(2)}_{ij}
        ={}& -\frac{1}{2\ell^{2}}\psi^{(2)}_{ij}
        - \frac{1}{2}\left( R^{(0)}_{ij} - \frac{1}{3}\gamma^{(0)}_{ij} R^{(0)} \right)
        + \frac{1}{8\ell^{2}}\gamma^{(1)}\psi^{(1)}_{ij}\,, \\
        \mathcal{E}^{(3)}_{ij}
        ={}& -\frac{3}{4\ell^{2}}\psi^{(3)}_{ij}
        - \frac{1}{12\ell^{2}}\gamma^{(0)}_{ij}\psi^{kl}_{(1)}\psi^{(2)}_{kl}
        - \frac{1}{16\ell^{2}}\psi^{(1)}_{ij}\psi^{(1)}_{kl}\psi^{kl}_{(1)} \\
        & + \frac{1}{12}\left( R^{(0)}\psi^{(1)}_{ij} - \gamma^{(0)}_{ij} R^{(0)}_{kl}\psi^{kl}_{(1)} + \gamma^{(0)}_{ij} D_l D_k \psi^{kl}_{(1)} \right. \\
        &\left.+ \frac{3}{2} D_k D^{k}\psi^{(1)}_{ij}
        - 3 D_k D_i \psi^{(1)k}{}_{j}\right)
        + \frac{1}{24\ell^{2}} E^{\gamma}_{ij}
        + i \leftrightarrow j~, 
    \end{aligned}
\end{equation}
where $\psi^{(n)}_{ij} = \gamma_{ij}^{(n)} -\tfrac13\gamma^{(0)}_{ij}\gamma^{(n)}$ is the trace-free part of the $n$-th coefficient in the FG expansion with $\gamma^{(n)}=\gamma_{(0)}^{ij}\gamma_{ij}^{(n)}$ being its trace, $D_i$ is the covariant derivative compatible with $\gamma^{(0)}$, and the following definition has been used
\begin{align}\notag
    E^{\gamma}_{ij}
        &:= \gamma_{(1)}\left( 3\psi^{(2)}_{ij} + \frac{1}{2}\gamma^{(0)}_{ij}\psi^{kl}_{(1)}\psi^{(1)}_{kl} - \gamma_{(1)}\psi^{(1)}_{ij} \right) + 5\gamma_{(2)}\psi^{(1)}_{ij}
        +  \ell^{2}\left( D_j D_i \gamma_{(1)} - \frac{1}{3}\gamma^{(0)}_{ij} D^{k}D_k \gamma_{(1)} \right)\,.
\end{align}
The holographic one-point functions were also obtained in Ref.~\cite{Grumiller:2013mxa}, and they are explicitly given by
\begin{align}
    \frac{1}{\alpha}P_{ij} &= -\frac{4}{\ell}{\cal E}^{(2)}_{ij}~, \\
\frac{1}{\alpha}\tau_{ij} &=-\frac{4}{\ell}{\cal E}^{(2)}_{ik}\psi^{(1)}_j{}^k+ \frac{2}{\ell}\left({\cal E}^{(3)}_{ij}+\frac13{\cal E}^{(2)}_{ij}\gamma^{(1)}\right) + \frac{1}{\ell}\gamma^{(0)}_{ij}{\cal E}^{(2)}_{kl}\psi^{kl}_{(1)} + \frac{1}{2\ell^3}\psi^{(1)}_{ij}\psi^{(1)}_{kl}\psi^{kl}_{(1)} \nonumber \\ &-\frac{1}{\ell^3}\psi^{(1)}_{kl}\left( \psi^{(1)k}_i\psi_j^{(1)l} - \frac13\gamma^{(0)}_{ij}\psi^{(1)k}_m\psi^{lm}_{(1)} \right) + 4D^k {\cal B}^{(1)}_{ijk} + i \leftrightarrow j~.
\end{align}

As part of the choice of boundary conditions, Ref.~\cite{Grumiller:2013mxa} fixed the leading- and next-to-leading-order terms in the FG expansion~\eqref{FG2} up to a local Weyl rescaling. This choice, in turn, is translated into a trace Ward identity for the holographic responses, that is,
\begin{align}\label{Pijtracepsi1}
    \gamma^{(0)}_{ij}P^{ij} = 0~,\qquad \gamma^{(0)}_{ij}\tau^{ij} + \frac12\psi^{(1)}_{ij}P^{ij} = 0~.
\end{align}
On the other hand, boundary diffeomorphisms generated by Killing vectors $\xi^i$ that preserve the asymptotic structure imply the Ward identity 
\begin{align}\label{Tijholo}
    D_iT^i{}_j = P^{ik}D_j\gamma^{(1)}_{ik}~, \quad \mbox{where} \quad T^i{}_j := 2\tau^i{}_j + 2P^{ik}\gamma^{(1)}_{jk}~,
\end{align}
is the modified boundary stress tensor. Additionally, the current
\begin{align}\label{Jicurrent}
    J^i[\xi] = T^i{}_j\xi^j~,
\end{align}
is conserved provided that $\pounds_\xi \gamma^{(0)}_{ij} = 0 = \pounds_\xi \gamma^{(1)}_{ij}$. This current induces conserved charges defined by
\begin{align}\label{HoloCharge}
    Q_{\rm hol}[\xi] = \int_{\cal C} i_J \varepsilon_{(0)} =\int_{\cal C}\dd^2 x \sqrt{q}\; s_i J^i[\xi]~,
\end{align}
where $i_J$ is the inner contraction along the current $J[\xi]$, $\cal C$ is an oriented two-dimensional hypersurface embedded in the conformal boundary, endowed with a metric $q_{AB}$, and $s_i$ is its unit normal within $\partial \cal M$.

For the solutions considered here, the FG gauge in Eq.~\eqref{FG-gauge} can be achieved by a redefinition of the radial coordinate $r=r(z)$ determined by
\begin{align}
    \frac{\dd r}{\dd z} = -\frac{\ell}{z}\sqrt{f(r)}~.
\end{align}
We perform this transformation for the general Bach-flat solution before imposing the NUT or bolt regularity conditions, which lets us identify the asymptotic data shared by both families and their dependence on the integration constants.
Solving this equation perturbatively near the conformal boundary, we obtain the asymptotic coordinate transformation
\begin{align}
    r(z) ={}& \frac{\ell^2}{z} - \frac{b\ell^2}{2} + \left[\frac{b^2\ell^3}{16}-\frac{\ell(1+a)}{4} + \frac{5n^2}{4\ell}  \right]\frac{z}{\ell} + \left[\frac{2m-bn^2}{6}\right]\frac{z^2}{\ell^2} \nonumber \\ &+\left[\frac{6b\ell^4m-3b^2\ell^4n^2 - 8a\ell^2n^2-12\ell^2n^2+48n^4}{48\ell^3}\right]\frac{z^3}{\ell^3}+{\cal O}(z^4)~. \label{radialdiff}
\end{align}

Because the diffeomorphism preserves the $SU(2)\times U(1)$ symmetry of the solution, the FG coefficients can be written as
\begin{align}\label{gammadecomp}
    \gamma_{ij}^{(k)}\dd x^i\dd x^j = A^{(k)}\left(\dd \tau + 2n\cos\theta\dd\phi\right)^2 + B^{(k)}\ell^2\left(\dd\theta^2+\sin^2\theta\dd\phi^2\right)~.
\end{align}
The coefficients $A^{(k)}$ and $B^{(k)}$ can be obtained by inspecting the metric asymptotically in the radial coordinate~\eqref{radialdiff}, giving 
\begin{align}\label{ABcoeff}
    A^{(0)} &= 1~, && B^{(0)} =1~, \\ A^{(1)} &= 0~, && B^{(1)} =-b\ell~, \\ 
    A^{(2)} &=\frac{1+a}{2}-\frac{b^2\ell^2}{8}-\frac{5n^2}{2\ell^2}~, &&B^{(2)}= -\frac{1+a}{2}+\frac{3b^2\ell^2}{8}+\frac{3n^2}{2\ell^2} ~, \\ A^{(3)} &= \frac{2(bn^2-2m)}{3\ell}~,&& B^{(3)} = \frac{2m}{3\ell}+\frac{b\ell(1+a)}{4}-\frac{b^3\ell^3}{16}-\frac{19bn^2}{12\ell}~.\label{ABcoeff2}
\end{align}
The leading order coefficients show that the conformal boundary of the Bach-flat Taub-NUT and Taub-Bolt instantons is a squashed three-sphere, viz;
\begin{align}
    \gamma^{(0)}_{ij}\dd x^i\dd x^j = \frac{L^2}{4}\left(\sigma_1^2 + \sigma_2^2 + \frac{\sigma_3^2}{1+\nu}\right)~, 
\end{align}
where we have defined the squashed-sphere radius and the squashing parameter
\begin{align}\label{nuparameter}
    L=2\ell~,\qquad \nu = \frac{\ell^2}{4n^2}-1~,
\end{align}
respectively. Additionally, $\sigma_a$ are the left-invariant forms of $SU(2)$ parametrized in terms of the Euler angles as
\begin{subequations}\label{LIF}
    \begin{align}
 \sigma_1 &= \cos\psi\dd\theta + \sin\theta\sin\psi\dd\phi~, \\
 \sigma_2 &= -\sin\psi\dd\theta + \sin\theta\cos\psi\dd\phi~, \\
 \sigma_3 &= \dd\psi + \cos\theta\dd\phi~,
\end{align}
\end{subequations}
with $\psi=\tau/2n$, obeying the Maurer-Cartan structure equations $\dd\sigma_a + \frac{1}{2}\epsilon_{abc}\sigma^b\wedge\sigma^c=0$. At this stage $\nu>-1$ and later we shall discuss the extreme-squashing limit $\nu\to-1$. 

The boundary is not conformally flat, since its Cotton-York tensor
\begin{align}\label{Cottondef}
    C^{(0)}_{ij} = \varepsilon_{i}{}^{kl}D_k\left(R_{jl} -\frac14 R \gamma^{(0)}_{jl}\right)~,
\end{align}
is nonvanishing; here, $\varepsilon_{ijk}$ is the codimension-1 Levi-Civita tensor. Indeed, its on-shell value for our family of solutions can be written using a fluid-like decomposition. As the squashed three-sphere can be described as a principal $U(1)$ bundle over $\mathbb{S}^2$, for our boundary geometry $\gamma^{(0)}$, we introduce
\begin{align}
    u := \dd\tau + 2n\cos\theta\dd\phi = 2n\sigma_3~,
\end{align}
as the dual one-form of the unit Killing vector field generating the Hopf fibers, where in the second equality the relation  $\psi=\tau/2n$ has been used. This object satisfies $u_i u^i = 1$ and $u^i\partial_i = \partial_\tau$. Therefore, its integral curves form a geodesic, shear-free, and expansion-free congruence. Then, its acceleration
\begin{align}
    a_i := u^j D_j u_i = -\frac12 D_i(u^2) = 0~,
\end{align}
and its expansion $\Theta := D_i u^i = 0$ as can be obtained directly by contracting the Killing equation with the boundary metric. Thus, the shear, $\sigma_{ij} = h_i{}^k h_j{}^l D_{(k}u_{l)}$, vanishes as well. Nevertheless, as we will show later, this does not imply that the vorticity is trivial. On the other hand, the orthogonal projector onto the two-dimensional base of the squashed three-sphere is defined as
\begin{align}
    h_{ij} := \gamma^{(0)}_{ij} - u_i u_j~.
\end{align}
The projector allows us to decompose the coefficients in the FG expansion~\eqref{FG2} as
\begin{align}
    \gamma^{(k)}_{ij}=A^{(k)}u_i u_j +B^{(k)}h_{ij}~,
\end{align}
with exactly the same values of $A^{(k)}$ and $B^{(k)}$ as given in \eqref{ABcoeff}-\eqref{ABcoeff2}. In this basis, the Cotton-York tensor for the Bach-flat Taub-NUT/Bolt instantons reads\footnote{Here, we have chosen the orientation ${\rm vol}_{(0)} = -u\wedge \ell^2 \sin\theta \dd\theta\wedge \dd\phi$.}
\begin{align}
    C^{(0)}_{ij} = \frac{n}{\ell^4}\left(1-\frac{4n^2}{\ell^2}\right)\left(h_{ij} - 2u_i u_j\right)~.
\end{align}
The boundary is conformally flat iff $n^2 = \frac{\ell^2}{4}$, corresponding to the round three-sphere. 

The holographic one-point functions of the general Bach-flat Taub-NUT/Bolt solution inherit the $SU(2)\times U(1)$ symmetry of the boundary geometry. Consequently, they can be expressed in a fluid-like fashion in terms of the unit fiber one-form $u_i$ and the projector $h_{ij}$. The first coefficients of the asymptotic expansion of the electric and magnetic components of the Weyl tensor are
\begin{align}
    {\cal E}^{(2)}_{ij} ={}& \frac{a}{6\ell^2}\left(h_{ij}-2u_iu_j\right)~, \nonumber \\ {\cal E}^{(3)}_{ij} ={}& \frac{2m-3bn^2}{2\ell^3}\left(2 u_i u_j - h_{ij}\right) - \frac{ab}{3\ell}u_iu_j~, \nonumber \\ {\cal B}^{(1)}_{ijk} ={}& \frac{b}{2}\left(u_k D_j u_i - u_j D_k u_i + 2u_i D_j u_k\right)~.
\end{align}
Thus, evaluating the holographic responses, we obtain
\begin{align}
    \tau_{ij} = \rho u_i u_j + p h_{ij}~,\qquad P_{ij} = {\cal P}(h_{ij} - 2u_i u_j)~,
\end{align}
where we have defined
\begin{align}
    \rho ={}& \frac{2\alpha}{3\ell^4}\left(ab\ell^2+18bn^2+12m\right)~, \\ p ={}& -\frac{2\alpha}{3\ell^4}\left(ab\ell^2+9bn^2+6m\right)~, \\ {\cal P}={}& -\frac{2\alpha a}{3\ell^3}~.
\end{align}
Although our analysis is performed in Euclidean signature, the modified holographic stress tensor in Eq.~\eqref{Tijholo} also admits the same algebraic decomposition as a stationary perfect fluid. The fiber congruence and the coefficients $u_i$, $\rho$, and $p$ appearing in this decomposition should therefore be understood as Euclidean analogs of the fluid velocity, energy density, and pressure, respectively. Unlike a three-dimensional conformal perfect fluid, the modified stress tensor is not generically traceless, owing to the coupling between the partially massless source and its response. A dynamical fluid/gravity interpretation of the corresponding Lorentzian solutions would require an independent investigation of perturbations and transport properties. Additionally, the partially massless response is traceless by construction, while
\begin{align}
    \tau^i{}_i = \rho + 2p = -\frac{2\alpha ab}{3\ell^2}~.
\end{align}
Since, $\psi^{(1)} = \frac{b\ell}{3}(2u_iu_j - h_{ij})$ we find $\psi^{(1)}_{ij}P^{ij} = \frac{4\alpha a b}{3\ell^2}$ and the trace Ward identity \eqref{Pijtracepsi1} is satisfied.

Having determined the holographic response tensors, we can now compute the conserved charges associated with the Killing vector $\xi = \partial_\tau$ using the modified holographic current \eqref{Jicurrent}. Since $\gamma^{(1)}_{i\tau} = 0$, the partially massless term does not contribute to this current, and hence $J^i[\partial_\tau] = 2\tau^i{}_\tau$. Moreover, the Killing vector points along the Hopf fiber such that
\begin{align}
    \tau^i{}_j\xi^j = \rho\, u^i u_j\xi^j + p\, h^i{}_j\xi^j~.
\end{align}
Then, using $ u_j\xi^j =1$ and $h^i{}_j\xi^j = 0$, the holographic conserved charge is proportional to $\rho$. Integrating over an angular section of the codimension-2 boundary, the ingredients of Eq.~\eqref{HoloCharge} are given by
\begin{align}
    \sqrt{q} = \ell\sqrt{\ell^2\sin^2\theta+4n^2\cos^2\theta}~,\qquad s_i \dd x^i = \frac{\ell\sin\theta}{\sqrt{\ell^2\sin^2\theta+4n^2\cos^2\theta}}\dd\tau~,
\end{align}
such that $\sqrt{q}s_i u^i = \ell^2\sin\theta$. Integrating over the constant-$\tau$ cuts yields\footnote{Strictly speaking, the angular integration is performed over a local section of the Hopf fibration, since the squashed three-sphere does not admit a globally defined constant-$\tau$ section. A global definition of the Euclidean conserved charge would require a careful treatment of the bundle structure.}
\begin{align}\label{HoloMass}
    Q_{\rm hol}[\partial_\tau] = 8\pi\ell^2\rho = \frac{16\pi \alpha}{3\ell^2}\left(ab\ell^2+18bn^2+12m\right)~.
\end{align}

The holographic charge differs from the Noether--Wald potential integral $Q_{\rm NW}$ of Ref.~\cite{Corral:2021xsu} by 
\begin{align}
    Q_{\rm hol}[\partial_\tau] - Q_{\rm NW}[\partial_\tau] = 64\pi\alpha bn^2/\ell^2 = -\frac{8\alpha b n}{\ell^2}\oint_{\cal C}\dd u~.
\end{align}
This difference is the integral of an exact form and proportional to the bundle's first Chern number, and it does not depend on the section. It depends only on boundary sources and vanishes at $b=0$ as the proportionality constant is given in terms of $\gamma^{(1)}_{ij}$. It would be interesting to see whether the extra divergence can be cured by an ambiguity in the covariant phase-space formalism, by background subtraction methods, or by adding topological invariants to the action. 

The on-shell action independently checks both responses as, in fixed Hopf coordinates, $(\partial_n I)|_{b,\ell} = 8\pi Q_{\rm hol}$ and $(\partial_b I)|_{n,\ell} = 128\pi^2\alpha n a/3$ along a differentiable regular branch. 
Since $\beta=8\pi n$, the first identity gives $Q_{\rm hol} = (\partial_\beta I)|_{\ell,b}$. 
This variation also changes the squashing, so it does not uniquely identify a Hamiltonian mass. In the covariant phase-space surface-charge one-form $\oint_{\cal C}(\delta Q - \xi\cdot\Theta)$, the flux term $-\xi\cdot\Theta$ vanishes at fixed sources, where the two charges have identical, integrable variations.
Indeed, for fixed sources
\begin{align}
    \oint_{\cal C}\delta Q_\xi  = \delta Q_{\rm hol}[\partial_\tau] = \delta Q_{\rm NW}[\partial_\tau] = \frac{16\pi\alpha}{3\ell^2}\left(b\ell^2\delta a + 12\delta m\right)~.
\end{align}
Both prescriptions therefore generate $\partial_\tau$ on the fixed-source phase space, and they differ by a constant on it. Allowing the source to vary can make this local one-form nonintegrable on the general Bach-flat family and the flux term $\xi\cdot\Theta$ itself does not select either absolute mass, and the obstruction corresponds to the work done by the partially massless source. 

Clearly, the modified holographic stress tensor~\eqref{Tijholo} also inherits the $SU(2)\times U(1)$ symmetry of the squashed three-sphere and admits a perfect-fluid-like decomposition, viz., 
\begin{align}
    T_{ij} = \bar\varepsilon\, u_i u_j + \bar{p}\,h_{ij}~,
\end{align}
with fiber energy density and transverse pressure given by
\begin{align}
\bar\varepsilon = \frac{4\alpha}{3\ell^4}(ab\ell^2+18bn^2+12m)~ \quad \mbox{and} \quad  \bar{p} = -\frac{4\alpha}{\ell^4}(3bn^2+2m)~,
\end{align}
respectively. Unlike a conformal perfect fluid, this tensor is not generically traceless,
\begin{align}
    T^i{}_i = \bar{\varepsilon} + 2\bar{p} = \frac{4\alpha ab}{3\ell^2}~,
\end{align}
which is controlled by the coupling between the partially massless source and its holographic response. 

As a separate check of our geometry, we recall that in Einstein gravity, the NUT charge can induce a fiducial angular momentum, whose value depends on the distribution of the Misner string along the polar axes; see, for instance, Refs.~\cite{Frodden:2021ces,Durka:2019ajz}. To explore this feature holographically in conformal gravity, we temporarily restore the parameter $C$ in the fiber one-form
\begin{align}\label{MisalignedOneFormFibre}
    u = \dd\tau + 2n(\cos\theta+C)\dd\phi~.
\end{align}
The holographic response tensors retain their covariant form, with the same scalar coefficients, while their coordinate components acquire an explicit dependence on $C$ through Eq.~\eqref{MisalignedOneFormFibre}. Integrating the angular current associated with the Killing vector $\xi = \partial_\phi$, we obtain
\begin{align}
    Q_{\rm hol}^{(C)}[\partial_\phi] = \frac{32\pi\alpha n C}{3\ell^2}\left(ab\ell^2 + 27bn^2+18m\right)~.
\end{align}
This expression vanishes in the symmetric convention $C=0$, consistent with the vanishing angular momentum obtained from the Noether--Wald formalism applied to the results of Ref.~\cite{Corral:2021xsu}, whereas asymmetric configurations yield a nonzero angular charge, as in the Einstein case studied in Ref.~\cite{Frodden:2021ces} and in the conformal gravity case in Ref.~\cite{Corral:2025npd}. Indeed, taking the rotation parameter to zero in the Bach-flat Kerr-NUT-AdS black hole, one recovers the same conserved charges as in~\cite{Corral:2025npd}. On the other hand, in the Einstein limit $a=b=0$, the expression reduces to $Q_{\rm hol}^{(C)}[\partial_\phi]=3CnQ_{\rm hol}^{(C)}[\partial_\tau]$, reproducing the same exact dependence found in Ref.~\cite{Frodden:2021ces}. In what follows, we return to the symmetric configuration $C=0$, with the Misner string distributed equally along the two polar axes.

Despite the perfect-fluid-like structure, the boundary configuration carries nontrivial vorticity, similarly to the holographic fluids associated with the Einstein Taub-NUT-AdS geometries~\cite{Caldarelli:2011idw,Leigh:2011au,Leigh:2012jv,Mukhopadhyay:2013gja,Kalamakis:2020aaj}. Indeed, we find
\begin{align}
    \omega_{ij}=h_i{}^k h_{j}{}^l D_{[k}u_{l]} = \frac{n}{\ell^2}\varepsilon^{(0)}_{ijk}u^k~,
\end{align}
and the total circulation is
\begin{align}
    {\cal C}_{\rm tot} = 2\int_{\mathbb{S}^2}\omega = -8\pi n~,
\end{align}
with $\omega=\tfrac12\omega_{ij}\dd x^i\wedge \dd x^j$. This result matches the vorticity flux in Einstein gravity (see, for instance, Ref.~\cite{Kalamakis:2020aaj}) and holds for NUTs and bolts, reflecting its dependence solely on the geometry of the common squashed three-sphere conformal boundary, independent of the bulk regularity conditions. Notice also that this result is independent of how the Misner string is distributed and is valid for any $C$, since the latter enters the fiber connection only through a closed contribution and therefore leaves its curvature, and consequently the total circulation, unchanged.

The nonvanishing vorticity also implies that $u\wedge \dd u\neq0$, so the distribution orthogonal to the fiber congruence is not hypersurface orthogonal. This reflects the nontrivial Hopf fibration and provides a geometric interpretation of the vortical configuration, analogous to that of holographic fluids with vorticity in Einstein gravity~\cite{Caldarelli:2011idw,Leigh:2011au,Leigh:2012jv,Mukhopadhyay:2013gja,Kalamakis:2020aaj}. Remarkably, these kinematical properties depend only on the leading FG metric and remain unchanged throughout the Bach-flat Taub family.

Thus, while the vorticity is determined entirely by the NUT charge, the one-point functions contain additional information about the non-Einstein modes. Furthermore, the common $U(1)$ fibration implies that globally regular scalar fluctuations obey the same angular quantization conditions as in Einstein Taub-NUT-AdS~\cite{Leigh:2011au,Kalamakis:2025daq}, independently of the radial metric function. Their radial spectrum, however, can distinguish the different Bach-flat branches.
Interestingly, the vortex quantization condition identified in Einstein Taub-NUT-AdS~\cite{Kalamakis:2025daq} is a universal feature of the $U(1)$ fibration rather than a consequence of the Einstein equations. Since its derivation relies solely on the global regularity and single-valuedness of scalar fluctuations, it extends naturally to the entire Bach-flat family, independently of the non-Einstein deformations encoded in $f(r)$.

\subsection{Partially massless condensation in bolt geometries}\label{SubSec:Condensate}

The additional asymptotic degrees of freedom of conformal gravity allow for a nonvanishing partially massless response even when its corresponding boundary source is switched off. The Bach-flat constraint of the Taub-NUT/Bolt family is given in Eq.~\eqref{BachFlatConstraint} and it can be solved by taking \begin{align}\label{condensatecondition}
    b= 0~,\qquad a = -2\left(1-\frac{4n^2}{\ell^2}\right)~,
\end{align}
such that $a$ vanishes only at $n^2 = \frac{\ell^2}{4}$ where the boundary becomes conformally flat. Then, the holographic response and its source become
\begin{align}\label{gamma1andPijcondensation}
    P_{ij} = \frac{4\alpha}{3\ell^3}\left(\frac{4n^2}{\ell^2}-1\right)\left(2u_iu_j - h_{ij}\right)~,\qquad \gamma^{(1)}_{ij} = 0~.
\end{align}
Thus, the partially massless operator develops a nonzero expectation value in the absence of its corresponding boundary source. We refer to this source-free one-point function as a \textit{partially massless condensate}. Here, condensation refers to the existence of a nonvanishing response at zero source, without implying a dynamical condensation mechanism or a spontaneous phase transition. 

A regular source-free NUT does not support a nontrivial partially massless condensate within our solution family, as the regularity conditions~\eqref{NUTregularityconds} force $a_{\rm NUT}$ to vanish for $b=0$. For the Bach-flat Taub-Bolt instanton, the situation is different, as its regularity conditions \eqref{boltregularityconds} do not impose a vanishing $a$ coefficient. 
Accordingly, this branch provides a candidate bulk realization of a nonvanishing partially massless condensate. Its existence must be restricted to the values of $n/\ell$ for which these regularity conditions admit a physical bolt radius $r_b>n$.

The holographic response also has a particularly simple dependence on the squashing parameter~\eqref{nuparameter} as
\begin{align}
    a = -\frac{2\nu}{1+\nu}\,.
\end{align}
Therefore, the partially massless condensate can be written as
\begin{align}
    P_{ij} = -\frac{4\alpha}{3\ell^3}\left(\frac{\nu}{1+\nu}\right)\left(2u_iu_j - h_{ij}\right)~,
\end{align}
which makes explicit the fact that, for the round-sphere point $\nu=0$, the holographic response vanishes. At this point, the boundary geometry becomes conformally flat, and its isometry group enhances to $SO(4)$. In contrast, for generic $\nu\neq 0$ (and $\nu>-1)$, the boundary has $SU(2)\times U(1)$ isometry and the response is non-vanishing. 
Notice also that, at the condensation point, the modified holographic stress tensor~\eqref{Tijholo} remains traceless since $b=0$, regardless of the nonvanishing response $P_{ij}$. This shows that the partially massless one-point function captures information not visible in the trace of the metric stress tensor.

Note that the condensate does not induce any additional symmetry breaking, as expected from the $SU(2)\times U(1)$ isometry of the Taub family. To see this explicitly, we express the response in terms of the left-invariant one-forms as
\begin{align}
    P_{ij}\dd x^i \dd x^j =\frac{4\alpha}{3\ell}\frac{\nu}{1+\nu}\left(\sigma_1^2+\sigma_2^2-\frac{2}{1+\nu}\sigma_3^2\right)~.
\end{align}
Under the left action of $SU(2)$, the forms $\sigma_a$ are invariant, while the residual $U(1)$ isometry rotates $\sigma_1$ and $\sigma_2$ into each other while leaving $\sigma_3$ unchanged. It follows that\footnote{This symmetry preservation is independent of the Misner string distribution. Indeed, restoring $C\neq0$ modifies the fiber one-form as in Eq.~\eqref{MisalignedOneFormFibre}, which is locally related to the symmetric configuration by shifting $\tau\to\tau + 2nC\phi$. The partially massless response retains its covariant form and remains invariant under the corresponding transformed $SU(2)\times U(1)$ Killing generators. Thus, redistributing the Misner string does not induce additional symmetry breaking.}
\begin{align}
    \pounds_\xi P_{ij} = 0~,\qquad \forall \xi \in \mathfrak{su}(2)\oplus \mathfrak{u}(1)~.
\end{align}

Symmetry-preserving condensates also arise in other quantum field theories. A familiar example is the fermion bilinear in Nambu--Jona--Lasinio-type models, where $\langle \bar\psi\psi\rangle \neq0$ at vanishing source preserves global isospin and baryon-number symmetries. Although the operators and dynamics are quite different, the analogy shows that a nonvanishing expectation value need not signal spontaneous breaking of the symmetries under consideration. Here, the partially massless response is a spin-two operator whose expectation value aligns with the anisotropy already present in the squashed boundary, and therefore preserves its full isometry group.

\begin{figure}
    \centering
    \includegraphics[width=0.6\linewidth]{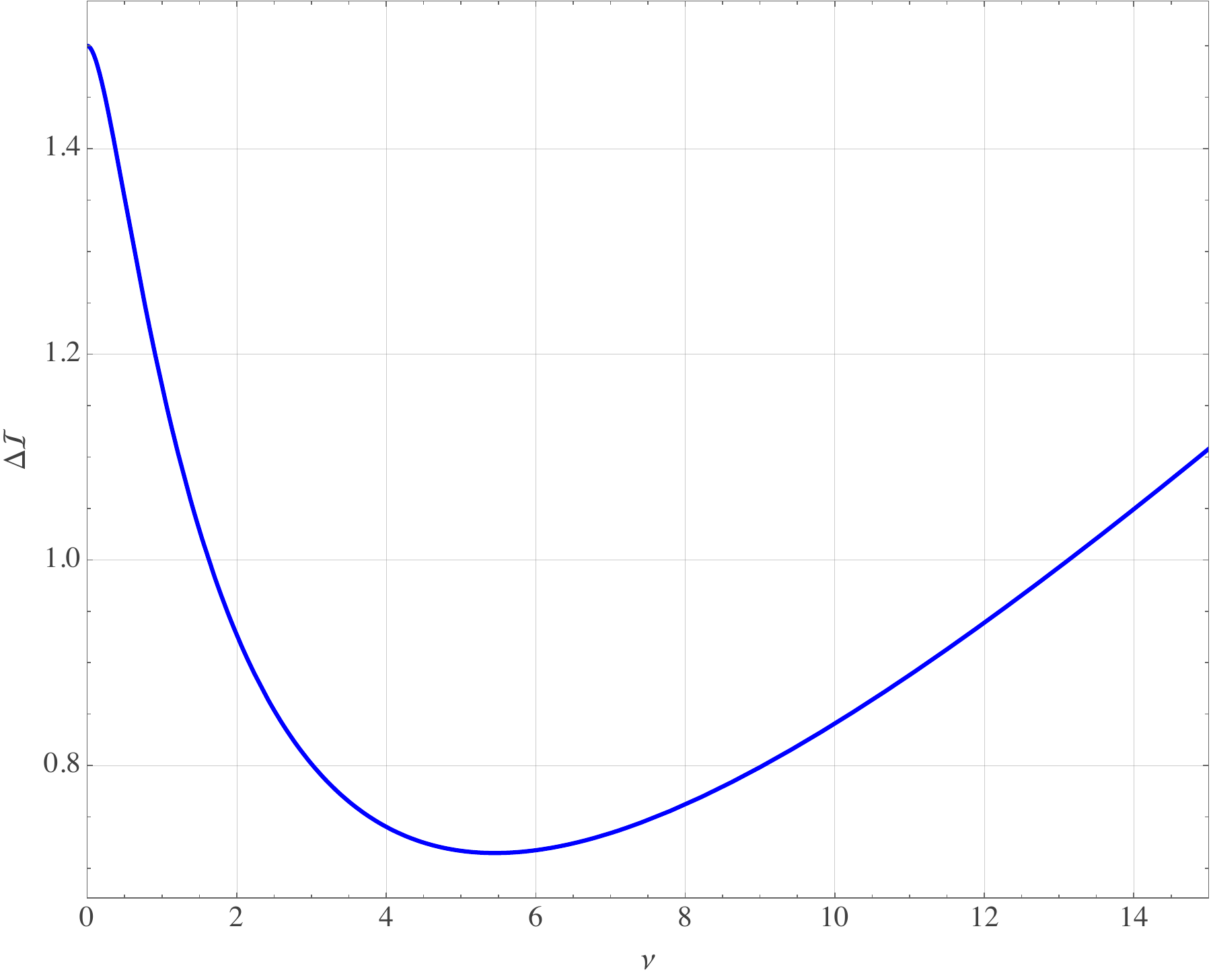}
    \caption{The difference ${\Delta {\cal I}} =  \frac{{I_{\mathrm{NUT}}^{\mathrm{Einstein}}-I_{\mathrm{bolt}}^{\mathrm{Cond}}}}{32\pi^2\alpha}$ between the normalized Euclidean on-shell actions of the Einstein Taub-NUT solution and the non-Einstein Taub-Bolt solution carrying the partially massless condensate, as a function of the boundary squashing parameter $\nu$. For thermally meaningful comparison, both geometries have the same Euclidean period $\beta=8\pi n$ and vanishing partially massless source $b=0$. Only regular bolt solutions with real $r_b>n$ are included. $\Delta{\cal I}$ remains positive throughout the admissible domain $\nu>0$, indicating that the bolt condensate is preferred over the Einstein NUT.}
    \label{fig:condensatedifference}
\end{figure}

To investigate the thermodynamic preference of the solution supporting the partially massless condensate, we compare the Euclidean on-shell action of the source-free non-Einstein Taub-Bolt instanton, characterized by the condition~\eqref{condensatecondition}, with that of the Einstein Taub-NUT instanton, for which $a=b=0$. Both geometries are required to have the same conformal boundary and Euclidean period, $\beta=8\pi n$. For the bolt, inserting this value of $\beta$ into the regularity condition~\eqref{boltbeta} reduces to a quartic equation for $y=r_b/n$, with one real solution satisfying $y>1$ that exists iff $\nu>0$. 
The difference between the NUT and bolt on-shell actions is shown in Fig.~\ref{fig:condensatedifference} as a function of the boundary squashing parameter $\nu$. The action difference remains strictly positive throughout this domain. Consequently, since $\alpha>0$, the non-Einstein Taub-Bolt carrying the partially massless condensate is thermodynamically preferred over the Einstein NUT. For $\nu > 5+3\sqrt{3}$, however, the Taub-Bolt, which shares the same boundary data, has an even lower action. 

The absence of a transition between the Taub-NUT and Taub-Bolt carrying the partially massless condensate stands in sharp contrast to the thermodynamic behaviour of the Taub-NUT/Bolt system in four-dimensional Einstein-AdS gravity. In the latter, a first-order Hawking-Page-type transition occurs when the large-bolt branch becomes thermodynamically preferred over the nut as the temperature increases~\cite{Chamblin:1998pz,Emparan:1999pm}. In our case, both on-shell actions are evaluated using the same Weyl-squared conformal-gravity action, and their difference remains strictly positive throughout the range where the condensate bolt is regular. Consequently, no analogous transition occurs between these two saddles. As we show in the next section, the thermodynamic structure becomes considerably richer when we consider the full space of Bach-flat solutions, allowing regimes with no phase transition, conventional first-order Hawking-Page-type transitions, and even reentrant phase transitions, depending on the parameters characterising the solutions.

\section{Thermodynamics and phase structure\label{sec:thermo}}

The Euclidean on-shell actions of the Bach-flat Taub-NUT and Taub-Bolt instantons in Eqs.~\eqref{Inut} and \eqref{Igenbolt}, respectively, serve as the starting point for investigating their thermodynamic phase structure in the semiclassical approximation. 
As we show, the additional deformation parameter $b$ significantly enriches the phase structure, including two new regimes: one in which the Hawking-Page-type transition is absent and another one with reentrant transitions. Furthermore, we show that regular non-Einstein NUT geometries preserve the scalar-sector stability Hartnoll and Kumar found for the Einstein counterpart.

\subsection{Reentrant and absent phase transitions}

The semiclassical gravitational partition function is constructed by summing over regular Euclidean saddles compatible with the prescribed boundary conditions.
Unlike Einstein gravity, the asymptotic boundary conditions of conformal gravity involve both $\gamma^{(0)}$ and the independent subleading source $\gamma^{(1)}$. 
From our holographic analysis, we found 
\begin{align}
    \gamma^{(0)}_{ij} = u_iu_j + h_{ij}~,\qquad \gamma^{(1)}_{ij} = -b\ell h_{ij}~.
\end{align}
Therefore, a thermally consistent comparison between nuts and bolts requires fixing both the squashing parameter $\nu$ and the non-Einstein source $b\ell$.

For both geometries to provide smooth fillings of the same squashed three-sphere boundary, they must share the Euclidean periodicity $\beta = 8\pi n$. While this condition is automatically satisfied by the Taub-NUT instanton, imposing it on the Taub-Bolt case, together with the Bach-flatness constraints and regularity conditions, leads to a sixth-degree algebraic equation for the bolt radius. Depending on the boundary parameters, this equation can admit several branches; we restrict our analysis to real solutions satisfying $r_b>n$ and $f(r)>0$, thereby retaining only admissible Euclidean bolt geometries. 

Within the semiclassical approximation, the dominant configuration is the regular saddle with the lowest Euclidean action. We compute the normalized difference as
\begin{align}
    \Delta {\cal I}(\nu,b) = \frac{I_{\rm NUT}(\nu,b) - I_{\rm Bolt}(\nu,b)}{32\pi^2\alpha}~,
\end{align}
such that $\Delta {\cal I} >0$ indicates that the bolt is preferred over the corresponding NUT, whereas $\Delta {\cal I}<0$ favors the NUT, since $\alpha>0$. 
\begin{figure}
    \centering
    \includegraphics[width=1\linewidth]{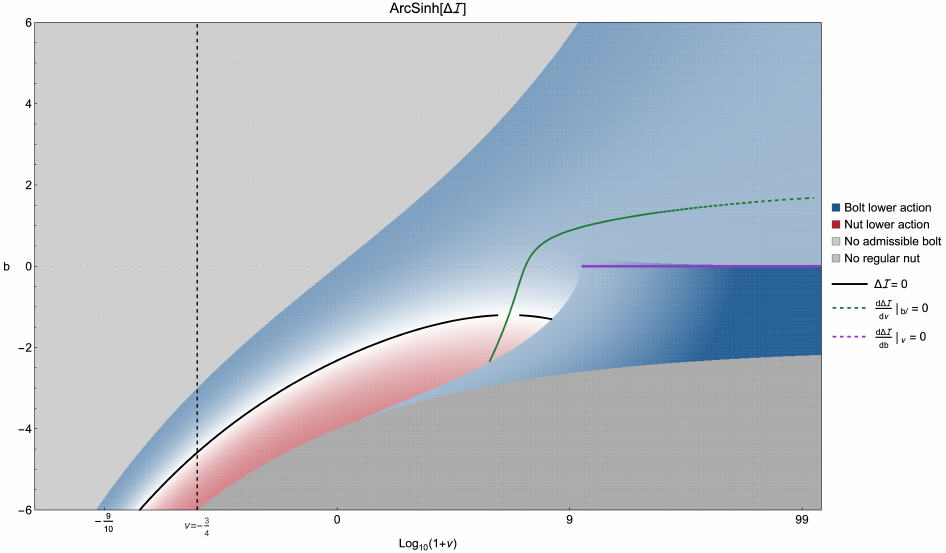}
    \caption{Thermodynamic phase diagram of the regular Bach-flat Taub-NUT and Taub-Bolt solutions in the $(\nu,b)$ parameter space with $\ell=1$. The color scale represents $\arcsinh\Delta{\cal I}$. The blue and red regions correspond to bolt and NUT dominance, respectively. The solid black curves indicate equality of the Euclidean on-shell actions, while the green and purple curves denote $\partial_\nu \Delta{\cal I} = 0$ at fixed $b$ and $\partial_b \Delta{\cal I} = 0$ at fixed $\nu$, respectively. Only regular bolt configurations with $r_b>n$ are included, and the lowest-action branch is selected whenever multiple admissible bolts coexist. Gray regions indicate the absence of a regular NUT or admissible bolt solution, as distinguished by the legend.}
    \label{fig:phasediagram}
\end{figure}
The phase diagram in Fig.~\ref{fig:phasediagram} shows that the Bach-flat deformation parameter $b$ strongly affects the instantons' thermodynamic behaviour. For $b\gtrsim-1.1$, the bolt configuration always has a lower Euclidean on-shell action than the admissible NUT case, and no NUT/Bolt phase transition occurs. For more negative values of $b$, the two geometries can exchange dominance as the squashing parameter varies, including regimes with reentrant phase transitions. 
Interestingly, the curves defined by the vanishing derivatives of the action difference do not intersect the action-equality curves within the numerical range considered. Therefore, the competing on-shell actions cross with different slopes, implying a discontinuity in the first derivatives of the dominant actions. This identifies the observed NUT/Bolt transitions as first-order. In particular, for suitable fixed values of the partially massless source, varying the squashing parameter leads to a Bolt-NUT-Bolt sequence, establishing the reentrant character of the phase structure. 
In principle, the appearance of a newly admissible bolt branch could also produce a zeroth-order transition if its action lies below that of all other admissible saddles.

An interesting comparison can be made with the large-$N$ $O(N)$ vector model on the squashed three-sphere studied in Ref.~\cite{Hartnoll:2005yc}. In contrast to the first-order NUT/Bolt transition of Einstein gravity, the strongly coupled vector model exhibits a smooth crossover between the corresponding regimes. 
Remarkably, our results show that the additional degrees of freedom of conformal gravity can also eliminate the conventional gravitational phase transition for a range of the partially massless source. On the other hand, Hartnoll and Kumar found that, in the strict weak-coupling limit, the vector model develops a phase transition at $\nu =-3/4$, where the scalar curvature of the boundary changes sign. Interestingly, our phase diagrams also allow a NUT/Bolt transition at precisely this value of the squashing parameter for a sufficiently negative partially massless source. Although the mechanisms are different, this provides another qualitative connection between the phase structure of the vector model and the Bach-flat gravitational saddles. 

It is also worth noting that no NUT-Bolt dominance switch appears along the $b=0$ line in Fig.~\ref{fig:phasediagram}, where the bolt dominates as soon as it exists $(\nu >0)$. At first sight, this seems surprising, since the Einstein Taub-NUT/Bolt system exhibits a first-order transition. However, setting $b=0$ together with $a=0$ does select the Einstein sector at the level of the geometry, but the conformal gravity action differs from the renormalized Einstein-AdS action by a contribution proportional to the Euler characteristic, which is nontrivial for both backgrounds. This produces a relative shift in their on-shell actions and consequently modifies their thermodynamic competition. 

The asymptotic behaviour of the Euclidean on-shell actions provides another interesting comparison with the strongly coupled $O(N)$ vector model. In the limit $\nu\to -1^+$, the NUT action behaves as $I_{\rm NUT} \sim -32\pi^2\alpha(1+\nu)^{-2}$, reproducing the leading power-law behaviour of Einstein gravity and the approximate strong-coupling scaling of the vector model, although the latter exhibits a logarithmic correction in a subsequent analytical treatment~\cite{Yonge:2006tn}. 
In the opposite limit, $\nu\to\infty$, the regular large-radius bolt branch without Misner strings behaves as $I_{\rm bolt}\sim -64\pi^2\alpha\nu/27$, reproducing the linear scaling of the strongly coupled vector-model free energy, as in Einstein gravity. Then, following the matching of coefficients as suggested in Ref.~\cite{Hartnoll:2005yc}, we find
\begin{align}
    \alpha = \frac{27\zeta(3)}{640\pi^4}N~.
\end{align}

Thus, conformal gravity reproduces the leading asymptotic behaviour of the strongly coupled $O(N)$ model in both extreme-squashing regimes, while allowing for a richer thermodynamic phase structure than Einstein gravity. 
In particular, the additional partially massless source introduces regions where the conventional NUT/Bolt transition disappears, in qualitative agreement with the smooth behaviour of the vector model. 
At the same time, sufficiently large deformations can produce reentrant transitions, illustrating how the additional degrees of freedom of conformal gravity modify the competition between gravitational saddles.
In the next subsection we investigate the stability of the regular NUT geometries under conformally coupled scalar perturbations, extending the analysis of Ref.~\cite{Hartnoll:2005yc} beyond the Einstein sector.

\subsection{Conformal scalar perturbations}

In Ref.~\cite{Hartnoll:2005yc}, Hartnoll and Kumar showed that the $O(N)$ model on the squashed three-sphere remains in a symmetric phase at finite coupling, even when the scalar curvature becomes negative. Motivated by this result, they investigated conformally coupled scalar fluctuations around the bulk dual Einstein Taub-NUT-AdS geometries and found no negative modes. We now extend this analysis to the regular non-Einstein Bach-flat Taub-NUT family\footnote{Applying our analysis to probes on Bolt backgrounds yields an inconclusive result due to the presence of a potential term arising from the Ricci scalar.}.

We consider a conformally coupled probe scalar field $\varphi$ governed by the Euclidean conformal Laplacian
\begin{align}
    {\cal L}_\varphi = -\nabla^2 + \frac16 R~,
\end{align}
and study the eigenvalue problem ${\cal L}_\varphi \varphi = \lambda \varphi$~. Negative eigenvalues $\lambda <0 $ would signal perturbative instabilities in the scalar sector. For the comparison with the critical $O(N)$ model, we adopt standard quantization, corresponding to a dual scalar operator of conformal dimension $\Delta=2$ with the scalar source set to zero.  

Let us now consider the regular Bach-flat NUT solution and introduce $x=r-n$ such that the metric function takes the form
\begin{align}
    f(n+x) = \frac{xQ(x)}{x+2n}~,\qquad Q(x) = 1+bx+\frac{x(x+4n)}{\ell^2}~.
\end{align}
Introducing the rescaled field $u=x\varphi$, the eigenvalue equation can be integrated against $u^\ast$ over the Euclidean geometry. After integration by parts, one obtains
\begin{align}\label{integratedconformallaplacian}
    \int_0^\infty \dd x\int\dd\tau\dd\Omega\Bigg[Q|\partial_xu|^2+&\frac{\left(x+2n\right)^2}{x^2 Q}|\partial_\tau u|^2 +\frac{1}{x^2\sin^2\theta}|(\partial_\phi - 2n\cos\theta\partial_\tau)u|^2  \nonumber \\ &+ \frac{1}{x^2}|\partial_\theta u|^2 \Bigg] = \lambda \int_0^\infty \dd x\int\dd\tau\dd\Omega\left(1+\frac{2n}{x}\right)|u|^2~,
\end{align}
provided that the radial boundary contribution vanishes at the NUT and at infinity, i.e.
\begin{align}
    -\int\dd\tau\dd\Omega \Big[u^\ast Q \partial_x u\Big]\Big\rvert_{0}^\infty = 0\,.
\end{align}
This can be checked at both endpoints by demanding regularity of the solution. Near the NUT $x=0$, regularity of the scalar implies $\varphi={\cal O}(1)$, and therefore $u={\cal O}(x)$. Since $Q(0)=1$, the boundary term satisfies
\begin{align}
    \lim_{x\to 0}u^\ast Q \partial_x u = 0~.
\end{align}
At conformal infinity, on the other hand, $Q(x)\sim x^2/\ell^2$. Let us assume a hypothetical negative eigenmode. Thus, the leading radial equation gives $u\sim x^{s_\pm}$, with
\begin{align}
    s_\pm = \frac{-1\pm\sqrt{1-4\lambda\ell^2}}{2}~.
\end{align}
Notice that, if $\lambda <0 $, one has $s_+>0$ and $s_-<-1$. The former is incompatible with the prescribed asymptotically AdS scalar boundary condition, leaving $u\sim x^{s_-}$. Consequently,
\begin{align}
    u^\ast Q \partial_x u \sim x^{2s_-+1}\quad {\to}\quad 0~,
\end{align}
at large $x$. Thus, the boundary terms vanish for any admissible negative eigenmode. 

Regularity of the Euclidean NUT geometry requires $Q(x)>0$ for $x>0$. Thus, every term on the left-hand side of the integral identity~\eqref{integratedconformallaplacian} is positive definite. On the other hand, the weight $1+2n/x$ appearing on the right-hand side is strictly positive for $x>0$. Therefore, if $\lambda <0$, the right-hand side would be negative for any nontrivial eigenfunction, contradicting the initial assumption. Hence $\lambda \geq 0$ for the admissible spectrum of the conformal Laplacian on regular Bach-flat Taub-NUT backgrounds. 

We conclude that the regular non-Einstein Bach-flat Taub-NUT geometries admit no negative modes of a conformally coupled probe scalar subject to standard quantization. Remarkably, this result holds despite the nonconstant bulk scalar curvature induced by the $b$ parameter, and remains valid when the curvature of the squashed conformal boundary is negative. It extends the scalar-sector stability of Ref.~\cite{Hartnoll:2005yc} beyond Einstein gravity, consistent with the persistence of the symmetric phase of the critical $O(N)$ model.

\subsection{Thermodynamic volume in conformal gravity}

An additional distinction between Einstein and conformal gravity arises in extended thermodynamics, where the effective cosmological constant is promoted to a thermodynamic variable~\cite{Kastor:2009wy} (see also~\cite{Kubiznak:2016qmn}). In conformal gravity, the AdS radius $\ell$ appears as an integration constant rather than a coupling of the action. Hence, it is natural to introduce an effective pressure~\cite{Xu:2018vxf}
\begin{align}
    P = \frac{3}{8\pi\ell^2}~,
\end{align}
while keeping the dimensionless coupling $\alpha$ fixed. 
The extended first law in the absence of additional varying boundary sources takes the form
\begin{align}
    \dd H = T\dd S + V\dd P~,
\end{align}
where $H$ is the enthalpy and $V$ its conjugate thermodynamic volume. Then, we identify $H$ with the conserved holographic charge associated to $\partial_\tau$, obtained in Eq.~\eqref{HoloMass}, and the corresponding entropy is thus determined by the Euclidean on-shell action through the Gibbs--Duhem relation
\begin{align}
    S = \beta H - I~.
\end{align}

To illustrate this distinction, consider the regular Taub-NUT solution in the Einstein regime $a=0=b$~. Although this geometry belongs to the Einstein sector, its thermodynamic properties are determined by the conformal gravity action~\eqref{action}.
Using the NUT on-shell action~\eqref{Inut} and the inverse temperature $\beta=8\pi n$, we obtain
\begin{align}\label{VnutEin}
    H_{\rm NUT}^{\rm Eins} = \frac{32\pi\alpha }{\ell}\frac{\nu}{(1+\nu)^\frac{3}{2}}~,\qquad V_{\rm NUT}^{\rm Eins}=\frac{128\pi^2\alpha \ell}{3}\frac{\nu}{(1+\nu)^\frac{3}{2}}~,
\end{align}
which satisfy the extended first law and the Smarr relation $H_{\rm NUT}^{\rm Eins} = 2PV_{\rm NUT}^{\rm Eins}$.
Remarkably, the volume changes sign at the round sphere point $\nu = 0$, being negative for $-1<\nu<0$ and positive for $\nu>0$. At the $\nu=0$ point, the geometry reduces to Euclidean AdS$_4$, whose Weyl tensor vanishes consistently giving $I =H_{\rm NUT}^{\rm Eins}=V_{\rm NUT}^{\rm Eins}=0$.

This contrasts with the Einstein Taub-NUT AdS, whose thermodynamic volume is always negative~\cite{Johnson:2014xza, Johnson:2014pwa}. 
This distinction arises from the different ensembles. In Einstein gravity, Newton's constant $G$ is held fixed, whereas in conformal gravity the fundamental coupling $\alpha$ remains fixed as the asymptotic curvature varies. 
Moreover, even within the Einstein sector, the conformal gravity action differs from the renormalised Einstein action by a finite contribution proportional to the Euler characteristic, which shifts the on-shell action and entropy (see for instance Refs.~\cite{Miskovic:2009bm,Anastasiou:2016jix}). 
Thus, in general, Einstein geometries do not need to have the same thermodynamic volume when embedded in conformal gravity.

The general Bach-flat Taub-NUT/Bolt solution allows for a nonvanishing partially massless source. This, in turn, sources the next-to-leading-order coefficient in the FG expansion whose variation contributes to the thermodynamic first law. Using the Euclidean on-shell action in Eq.~\eqref{Inut} and the holographic charge \eqref{HoloMass} we find that the enthalpy and the entropy are
\begin{align}
    H_{\rm NUT} &= \frac{32\pi\alpha}{\ell\sqrt{1+\nu}}\left[\frac{\nu}{1+\nu} -\frac{b^2\ell^2}{4}\right]~, \\
    S_{\rm NUT} &= \beta H_{\rm NUT} - I_{\rm NUT} = 32\pi^2\alpha\left[ \frac{\nu(\nu+4)}{(1+\nu)^2}-\frac{b^2\ell^2}{2(1+\nu)} \right]~,
\end{align}
respectively. Keeping $\alpha$ fixed, the exact variation of these quantities satisfies the extended first law, that is,
\begin{align}
    \dd H_{\rm NUT} = T\dd S_{\rm NUT} + V_{\rm NUT} \dd P + \Psi_b \dd b + \Psi_\nu\dd\nu~,
\end{align}
where the conjugate variables to the volume, partially massless response and boundary squashing parameter are
\begin{align}
    V_{\rm NUT} &={} \frac{128\pi^2\alpha\ell \nu}{3({1+\nu})^{\frac32}}~, & \Psi_b &={} -\frac{8\pi\alpha b\ell}{\sqrt{1+\nu}}~, & \Psi_\nu &={} 0~,
\end{align}
respectively. For the holographic choice, $\Psi_\nu = 0$ through the NUT family, while $\Psi_b$ vanishes at $b=0$. The volume $V_{\rm NUT}$ equals its value \eqref{VnutEin} in the Einstein sector and is computed by varying the pressure at fixed $b$. 
Furthermore, now in the presence of the new work terms, the Smarr relation modifies to
\begin{align}
    H_{\rm NUT} = 2PV_{\rm NUT} + b\Psi_b~,
\end{align}
and note that
\begin{align}
\left.\frac{\partial I_{\rm NUT}}{\partial n}\right|_{b,\ell} = 8\pi Q_{\rm hol}[\partial_\tau]~,
\end{align}
which checks the response-based choice $H_{\rm NUT} = Q_{\rm hol}[\partial_\tau]$. Since varying $n$ also changes the boundary squashing, it does not by itself select a unique Hamiltonian mass.  

Finally, let us analyze the condensate studied in Sec.~\ref{SubSec:Condensate}. Using the holographic mass and the Euclidean on-shell action, we verify that the extended first law and Smarr relation hold on this branch without the additional work terms and
\begin{align}
    V_{\rm cond} = \frac{32\pi^2\alpha\ell}{3\sqrt{1+\nu}}\frac{(y^2-1)(3y^4-12y^3+6y^2-4y-1)}{(3y^2+1)^2}~,
\end{align}
where $y=r_b/n$ has been defined. Remarkably, the thermodynamic volume can be negative despite the bolt's nonvanishing area. It vanishes at the critical squashing parameter $\nu_\ast \simeq 2.74$, such that $V_{\rm cond}<0$ for $\nu < \nu_\ast$ and $V_{\rm cond}>0$ for $\nu>\nu_\ast$.

Thus, negative thermodynamic volumes in conformal gravity are not restricted to NUT geometries, but can also arise in non-Einstein bolt configurations. The fully general Bach-flat Taub-Bolt instanton with $b\neq0$ requires additional work terms associated with variations of the partially massless source and boundary geometry, leading to a more involved thermodynamic phase space that we do not pursue here. 
 
\section{Discussion\label{sec:discussion}}

In this work, we investigate the holographic properties and thermodynamics of Bach-flat Taub-NUT/Bolt instantons in four-dimensional conformal gravity. We determine their asymptotic data and holographic one-point functions under weakened AdS boundary conditions, showing that the boundary stress tensor admits a perfect fluid-like structure with nonvanishing vorticity. 
The additional partially massless sector modifies the holographic Ward identities and contributes to the conserved charges. 
This yields a non-Einstein Taub-Bolt branch supporting a source-free partially massless condensate that preserves the full symmetries of the squashed three-sphere. 

We also show that the additional gravitational degrees of freedom significantly enrich the thermodynamic phase structure. 
Depending on the partially massless source, the NUT/Bolt system exhibits conventional, absent, and reentrant first-order phase transitions. In particular, the transition disappears in certain regions of the parameter space, resembling the smooth crossover of the strongly coupled $O(N)$ vector model studied in Ref.~\cite{Hartnoll:2005yc}. We recover the leading asymptotic scaling of its free energy in the extreme-squashing regimes. 
Furthermore, we establish the absence of negative modes for a conformally coupled probe scalar on regular Bach-flat NUT backgrounds, including the case of negative boundary scalar curvature. 

Furthermore, we investigate the extended thermodynamics in which the AdS radius appears as an integration constant rather than a coupling of the theory. We found that the thermodynamic volume can have either sign or be trivial, including the Einstein sector of conformal gravity.
Somewhat surprisingly, the non-Einstein Taub-Bolt configurations supporting the partially massless condensate can also have negative thermodynamic volume, despite their nonvanishing bolt area.

Finally, the perfect-fluid-like structure of the holographic stress tensor suggests exploring a fluid/gravity correspondence in conformal gravity, where the reentrant transitions could correspond to transitions between distinct vortical equilibrium states of the dual three-dimensional theory. Establishing this interpretation would require studying dynamical Bach-flat perturbations and determining the corresponding transport coefficients, including contributions from the partially massless sector. 

Among the questions our work leaves open, one concerns whether the one-loop corrections to the conformal gravity partition function studied in Ref.~\cite{Irakleidou:2015exd}, could account for the logarithmic behaviour of the free energy of the strongly coupled $O(N)$ vector model found in Ref.~\cite{Yonge:2006tn}. 
A natural setting for this investigation is the extreme-squashing limit $\nu\to-1^+$, where the NUT geometry approaches Bergman space (see Appendix \ref{App:A}), whose holographic properties were studied in Ref.~\cite{Britto-Pacumio:1999dwj}.
Another direction to pursue is to extend the holographic two-point function analysis of Ref.~\cite{Ghodsi:2014hua} by linearly perturbing the Bach-flat Taub-NUT/Bolt instantons studied here, providing a way to explore the spectrum and possible instabilities of the dual theory.

The soft gravitational physics associated with NUT charges provides another promising direction, particularly in view of their connections to soft theorems and gravitational memory effects~ \cite{Godazgar:2019dkh,Crawley:2023brz,Doran:2026bng}.
Recently, the conformal $w$-algebra was proposed as an extension of the soft symmetry algebra of Einstein gravity in flat an (A)dS backgrounds~\cite{Heuveline:2026nxq}. Understanding whether these symmetries admit a realisation on Bach-flat Taub-NUT geometries, and how the partially massless sector contributes to the associated charges, could shed light on the interplay between NUT charges and soft gravitational physics in conformal gravity.

Finally, the comparison with the $O(N)$ vector model is motivated by its conjectured holographic duality with Vasiliev's higher-spin gravity~\cite{Vasiliev:1990en,Vasiliev:1992av, Vasiliev:1999ba,Vasiliev:2003ev}.
An important open question in higher-spin gravity is whether a phase with broken higher-spin symmetry and a mass gap can arise within a large string-theoretic framework dual to a matrix model rather than a vector model. 
In this context, our results raise the possibility that conformal gravity, rather than Einstein gravity alone, could play a role in the effective description of such a phase. Higher-spin counterparts of the Taub-NUT/Bolt instantons provide a natural setting for exploring this idea. Indeed, higher-spin generalizations of Type D gravitational instantons have been constructed in Euclidean chiral models, and higher-spin extensions of Taub-NUT geometries have also been investigated~\cite{Iazeolla:2007wt,Iazeolla:2011cb,Skvortsov:2025ohi}. Their holographic properties could be studied within the framework of Refs.~\cite{Diaz:2024iuz,Diaz:2024kpr}, where, using the proposal of Ref.~\cite{Vasiliev:2012vf} to circumvent the Maldacena--Zhiboedov theorem \cite{Maldacena:2011jn}, four-dimensional Vasiliev gravity and three-dimensional conformal matter coupled to conformal higher-spin gauge fields arise as reductions of a common Frobenius–Chern–Simons parent theory~\cite{Bonezzi:2016ttk,Boulanger:2015kfa}. The proposed correspondence, formulated through overlap conditions on invariant functionals, could provide a setting for comparing higher-spin gravitational saddles with the thermodynamics of the gauged conformal theory on the squashed three-sphere.

\acknowledgments

The authors thank Giorgos Anastasiou, Brenno Vallilo, Nikolay Bobev, Nicolas Boulanger, Gastón Giribet, Carlo Iazeolla, Iva Lovrekovic, Robert B. Mann, Ayan Mukhopadhyay, Rodrigo Olea, Julio Oliva, Francisco Silva, Kostas Skenderis, and Ricardo Troncoso for valuable comments and suggestions. This work is partially supported by Agencia Nacional de Investigación y Desarrollo (ANID), Chile, through Fondecyt grants No.~1240043, 1240048, 1250672,  1252053, 1261016, and 3260880. The work of PS is also supported by the MATH-AMSUD project SGP 24-MATH-12 funded by ANID and Minist\`ere de l'Europe et des Affaires \`Etrang\`eres.

\begin{appendix}
\section{Bergman limit}\label{App:A}
In this appendix, we examine the extreme-squashing limit, $\nu \to -1^+$ (which corresponds to $n\to\infty$), of the Taub-NUT solution and show that its leading geometry approaches the Bergman metric on the complex hyperbolic plane. 

Let us now introduce the dimensionless radial coordinate 
\begin{align}
    u=\frac{n(r -n)}{\ell^2}~.
\end{align}
The nondegenerate limiting geometry is obtained by taking $n\to\infty$ while keeping $u$ fixed. Under this rescaling, the metric function \eqref{f-nut} becomes
\begin{align}
    f_{\rm NUT}=\frac{\ell^2u}{2n^2+\ell^2u}\left[1+4u+\frac{b\ell^2}{n}u+\frac{\ell^2}{n^2}u^2\right]~.
\end{align}
For fixed $b$, the leading behaviour 
\begin{align}
    f_{\rm NUT}=\frac{\ell^2}{2n^2}u(1+4u)+{\cal O}\left(\frac{\ell^3}{n^3}\right)~,
\end{align}
does not depend on the non-Einstein mode $b$.
Thus, the limiting metric reads
\begin{align}
    \dd s^2_{\rm lim}=2\ell^2\left[\frac{\dd u^2}{u(1+4u)}+u(\sigma_1^2+\sigma_2^2)+u(1+4u)\sigma_3^2\right]~.
\end{align}
To identify this geometry, we introduce a new radial coordinate through
\begin{align}
    u = \frac{\rho^2}{4(1-\rho^2)}~,\qquad 0\leq \rho<1~,
\end{align}
such that
\begin{align}
    \dd s^2_{\rm lim}=\frac{\ell^2}{2}\left[\frac{4\dd\rho^2}{(1-\rho^2)^2}+\frac{\rho^2}{1-\rho^2}(\sigma_1^2+\sigma_2^2)+\frac{\rho^2}{(1-\rho^2)^2}\sigma_3^2\right]~.
\end{align}
This is the Bergman metric on the complex hyperbolic plane, $\mathbb{CH}^2=SU(2,1)/U(2)$, expressed in coordinates adapted to the Hopf fibration. 

The identification can also be made directly from its K\"ahler description (see for instance Ref.~\cite{Parker}). Consider the unit ball in $\mathbb{C}^2$, defined by $|z_1|^2+|z_2|^2 < 1$, and with K\"ahler potential
\begin{align}
    K = -\log(1-|z_1|^2-|z_2|^2)~.
\end{align}
Here $z^i$, with $i=1,2$, denote complex coordinates, while barred indices label their antiholomorphic counterparts. The associated Hermitian metric is
\begin{align}
g_{i\bar j}=\partial_i\partial_{\bar j}K=\frac{\delta_{ij}}{1-|z|^2}+\frac{\bar z_i z_j}{(1-|z|^2)^2}~.
\end{align}
Then, using the parametrization
\begin{align}
    z_1=\rho\cos\frac{\theta}{2}e^{i(\psi+\phi)/2},\qquad z_2=\rho\sin\frac{\theta}{2}e^{i(\psi-\phi)/2}~,
\end{align}
the real line element $\dd s^2=2\ell^2g_{i\bar j}\dd z^i \dd \bar z^j$ reproduces exactly the limiting metric above. This establishes the equivalence between the conventional description in terms of the logarithm of the Bergman kernel and the biaxial Bianchi IX form obtained from Taub-NUT.

We note that the resulting spacetime is Einstein, but is not conformally flat, with conformal boundary metric
\begin{align}
    \lim_{\rho\to1^-}(1-\rho^2)^2\dd s^2_{\rm lim}\big|_{\rho=\mathrm{const}}=\frac{\ell^2}{2}\sigma_3^2~,
\end{align}
which is a degenerate rank one tensor. Hence, the limit does not define an ordinary nondegenerate three-dimensional conformal boundary metric, but rather it carries the natural contact $H={\rm ker}\,\sigma_3$ and Cauchy-Riemann structures associated with the complex hyperbolic ball. The extreme-squashing limit therefore changes the asymptotic geometry underlying the usual Fefferman--Graham expansion; its holographic aspects have been studied in Ref.~\cite{Britto-Pacumio:1999dwj}. This provides a concrete way in which Bergman geometry emerges from a continuous family of asymptotically locally AdS solutions, including non-Einstein configurations. It would be interesting to investigate whether the conformal gravity partially massless sector admits a meaningful boundary description in terms of the Cauchy-Riemann data of $\mathbb{S}^3$. 
    
\end{appendix}

\bibliographystyle{JHEP}
\bibliography{References.bib}

\providecommand{\href}[2]{#2}\begingroup\raggedright\begin{thebibliography}{100}

\bibitem{Adler:1982ri}
S.L.~Adler, \emph{{Einstein Gravity as a Symmetry-Breaking Effect in Quantum
  Field Theory}}, \href{https://doi.org/10.1103/RevModPhys.54.729}{\emph{Rev.
  Mod. Phys.} {\bfseries 54} (1982) 729}.

\bibitem{Stelle:1976gc}
K.S.~Stelle, \emph{{Renormalization of Higher Derivative Quantum Gravity}},
  \href{https://doi.org/10.1103/PhysRevD.16.953}{\emph{Phys. Rev. D} {\bfseries
  16} (1977) 953}.

\bibitem{Stelle:1977ry}
K.S.~Stelle, \emph{{Classical Gravity with Higher Derivatives}},
  \href{https://doi.org/10.1007/BF00760427}{\emph{Gen. Rel. Grav.} {\bfseries
  9} (1978) 353}.

\bibitem{Buchbinder:1984ebp}
I.L.~Buchbinder, \emph{{Renormalization group equations in curved space-time}},
  \href{https://doi.org/10.1007/BF01035006}{\emph{Theor. Math. Phys.}
  {\bfseries 61} (1984) 1215}.

\bibitem{Buchbinder:2026bfn}
I.L.~Buchbinder, P.M.~Lavrov, T.M.~Sangy and I.L.~Shapiro, \emph{{On the
  Renormalization in Conformal Quantum Gravity}},
  \href{https://arxiv.org/abs/2608.09858}{{\ttfamily 2608.09858}}.

\bibitem{Witten:2003nn}
E.~Witten, \emph{{Perturbative gauge theory as a string theory in twistor
  space}}, \href{https://doi.org/10.1007/s00220-004-1187-3}{\emph{Commun. Math.
  Phys.} {\bfseries 252} (2004) 189}
  [\href{https://arxiv.org/abs/hep-th/0312171}{{\ttfamily hep-th/0312171}}].

\bibitem{Berkovits:2004hg}
N.~Berkovits, \emph{{An Alternative string theory in twistor space for N=4
  superYang-Mills}},
  \href{https://doi.org/10.1103/PhysRevLett.93.011601}{\emph{Phys. Rev. Lett.}
  {\bfseries 93} (2004) 011601}
  [\href{https://arxiv.org/abs/hep-th/0402045}{{\ttfamily hep-th/0402045}}].

\bibitem{Berkovits:2004jj}
N.~Berkovits and E.~Witten, \emph{{Conformal supergravity in twistor-string
  theory}}, \href{https://doi.org/10.1088/1126-6708/2004/08/009}{\emph{JHEP}
  {\bfseries 08} (2004) 009}
  [\href{https://arxiv.org/abs/hep-th/0406051}{{\ttfamily hep-th/0406051}}].

\bibitem{Maldacena:2011mk}
J.~Maldacena, \emph{{Einstein Gravity from Conformal Gravity}},
  \href{https://arxiv.org/abs/1105.5632}{{\ttfamily 1105.5632}}.

\bibitem{Anastasiou:2016jix}
G.~Anastasiou and R.~Olea, \emph{{From conformal to Einstein Gravity}},
  \href{https://doi.org/10.1103/PhysRevD.94.086008}{\emph{Phys. Rev. D}
  {\bfseries 94} (2016) 086008}
  [\href{https://arxiv.org/abs/1608.07826}{{\ttfamily 1608.07826}}].

\bibitem{Anastasiou:2020mik}
G.~Anastasiou, I.J.~Araya and R.~Olea, \emph{{Einstein Gravity from Conformal
  Gravity in 6D}}, \href{https://doi.org/10.1007/JHEP01(2021)134}{\emph{JHEP}
  {\bfseries 01} (2021) 134}
  [\href{https://arxiv.org/abs/2010.15146}{{\ttfamily 2010.15146}}].

\bibitem{Anastasiou:2026jrt}
G.~Anastasiou, I.J.~Araya, N.~Boulanger, R.~Olea and D.~Rovere,
  \emph{{Conformal Renormalisation of 8D Einstein Gravity}},
  \href{https://arxiv.org/abs/2607.03679}{{\ttfamily 2607.03679}}.

\bibitem{Lu:2011ks}
H.~Lu, Y.~Pang and C.N.~Pope, \emph{{Conformal Gravity and Extensions of
  Critical Gravity}},
  \href{https://doi.org/10.1103/PhysRevD.84.064001}{\emph{Phys. Rev. D}
  {\bfseries 84} (2011) 064001}
  [\href{https://arxiv.org/abs/1106.4657}{{\ttfamily 1106.4657}}].

\bibitem{Hell:2023rbf}
A.~Hell, D.~Lust and G.~Zoupanos, \emph{{On the ghost problem of conformal
  gravity}}, \href{https://doi.org/10.1007/JHEP08(2023)168}{\emph{JHEP}
  {\bfseries 08} (2023) 168}
  [\href{https://arxiv.org/abs/2306.13714}{{\ttfamily 2306.13714}}].

\bibitem{Maldacena:1997re}
J.M.~Maldacena, \emph{{The Large $N$ limit of superconformal field theories and
  supergravity}}, \href{https://doi.org/10.4310/ATMP.1998.v2.n2.a1}{\emph{Adv.
  Theor. Math. Phys.} {\bfseries 2} (1998) 231}
  [\href{https://arxiv.org/abs/hep-th/9711200}{{\ttfamily hep-th/9711200}}].

\bibitem{Witten:1998qj}
E.~Witten, \emph{{Anti de Sitter space and holography}},
  \href{https://doi.org/10.4310/ATMP.1998.v2.n2.a2}{\emph{Adv. Theor. Math.
  Phys.} {\bfseries 2} (1998) 253}
  [\href{https://arxiv.org/abs/hep-th/9802150}{{\ttfamily hep-th/9802150}}].

\bibitem{Gubser:1998bc}
S.S.~Gubser, I.R.~Klebanov and A.M.~Polyakov, \emph{{Gauge theory correlators
  from noncritical string theory}},
  \href{https://doi.org/10.1016/S0370-2693(98)00377-3}{\emph{Phys. Lett. B}
  {\bfseries 428} (1998) 105}
  [\href{https://arxiv.org/abs/hep-th/9802109}{{\ttfamily hep-th/9802109}}].

\bibitem{Grumiller:2013mxa}
D.~Grumiller, M.~Irakleidou, I.~Lovrekovic and R.~McNees, \emph{{Conformal
  gravity holography in four dimensions}},
  \href{https://doi.org/10.1103/PhysRevLett.112.111102}{\emph{Phys. Rev. Lett.}
  {\bfseries 112} (2014) 111102}
  [\href{https://arxiv.org/abs/1310.0819}{{\ttfamily 1310.0819}}].

\bibitem{Riegert:1984hf}
R.J.~Riegert, \emph{{The particle content of linearized conformal gravity}},
  \href{https://doi.org/10.1016/0375-9601(84)90648-0}{\emph{Phys. Lett. A}
  {\bfseries 105} (1984) 110}.

\bibitem{Lu:2011zk}
H.~Lu and C.N.~Pope, \emph{{Critical Gravity in Four Dimensions}},
  \href{https://doi.org/10.1103/PhysRevLett.106.181302}{\emph{Phys. Rev. Lett.}
  {\bfseries 106} (2011) 181302}
  [\href{https://arxiv.org/abs/1101.1971}{{\ttfamily 1101.1971}}].

\bibitem{Deser:2012euu}
S.~Deser, E.~Joung and A.~Waldron, \emph{{Gravitational- and Self- Coupling of
  Partially Massless Spin 2}},
  \href{https://doi.org/10.1103/PhysRevD.86.104004}{\emph{Phys. Rev. D}
  {\bfseries 86} (2012) 104004}
  [\href{https://arxiv.org/abs/1301.4181}{{\ttfamily 1301.4181}}].

\bibitem{Deser:2012qg}
S.~Deser, E.~Joung and A.~Waldron, \emph{{Partial Masslessness and Conformal
  Gravity}}, \href{https://doi.org/10.1088/1751-8113/46/21/214019}{\emph{J.
  Phys. A} {\bfseries 46} (2013) 214019}
  [\href{https://arxiv.org/abs/1208.1307}{{\ttfamily 1208.1307}}].

\bibitem{Boulanger:2023lgd}
N.~Boulanger, G.~Lhost and S.~Thom{\'e}e, \emph{{Consistent Couplings between a
  Massive Spin-3/2 Field and a Partially Massless Spin-2 Field}},
  \href{https://doi.org/10.3390/universe9110482}{\emph{Universe} {\bfseries 9}
  (2023) 482} [\href{https://arxiv.org/abs/2310.05522}{{\ttfamily
  2310.05522}}].

\bibitem{Boulanger:2024qrn}
N.~Boulanger, G.~Lhost and S.~Thom{\'e}e, \emph{{A note on partially massless
  supergravity}}, \href{https://doi.org/10.1007/JHEP07(2025)185}{\emph{JHEP}
  {\bfseries 07} (2025) 185}
  [\href{https://arxiv.org/abs/2412.17713}{{\ttfamily 2412.17713}}].

\bibitem{Chen:2018dqz}
Q.~Chen and Y.~Ma, \emph{{Hamiltonian structure and connection-dynamics of Weyl
  gravity}}, \href{https://doi.org/10.1103/PhysRevD.98.064009}{\emph{Phys. Rev.
  D} {\bfseries 98} (2018) 064009}
  [\href{https://arxiv.org/abs/1803.10807}{{\ttfamily 1803.10807}}].

\bibitem{Dolan:2001ih}
L.~Dolan, C.R.~Nappi and E.~Witten, \emph{{Conformal operators for partially
  massless states}},
  \href{https://doi.org/10.1088/1126-6708/2001/10/016}{\emph{JHEP} {\bfseries
  10} (2001) 016} [\href{https://arxiv.org/abs/hep-th/0109096}{{\ttfamily
  hep-th/0109096}}].

\bibitem{Guajardo:2026vzh}
L.~Guajardo, L.G.~Frez, S.N.~Sajadi and J.~Oliva, \emph{{Black holes and phase
  transitions in Conformal Gravity}},
  \href{https://arxiv.org/abs/2609.15737}{{\ttfamily 2609.15737}}.

\bibitem{Mannheim:1990ya}
P.D.~Mannheim and D.~Kazanas, \emph{{Solutions to the Kerr and Kerr-Newman
  problems in fourth order conformal Weyl gravity}},
  \href{https://doi.org/10.1103/PhysRevD.44.417}{\emph{Phys. Rev. D} {\bfseries
  44} (1991) 417}.

\bibitem{Asuncion:2025cxw}
M.Y.~Asuncion, K.~Horne, R.~Kusano and M.~Dominik, \emph{{Structure of Kerr
  black hole spacetimes in Weyl conformal gravity}},
  \href{https://doi.org/10.1088/1361-6382/ae1109}{\emph{Class. Quant. Grav.}
  {\bfseries 42} (2025) 215005}
  [\href{https://arxiv.org/abs/2507.01379}{{\ttfamily 2507.01379}}].

\bibitem{Riegert:1984zz}
R.J.~Riegert, \emph{{Birkhoff's Theorem in Conformal Gravity}},
  \href{https://doi.org/10.1103/PhysRevLett.53.315}{\emph{Phys. Rev. Lett.}
  {\bfseries 53} (1984) 315}.

\bibitem{Schmidt:1999vp}
H.-J.~Schmidt, \emph{{A New conformal duality of spherically symmetric
  space-times}}, {\emph{Annalen Phys.} {\bfseries 9SI} (2000) 158}
  [\href{https://arxiv.org/abs/gr-qc/9905103}{{\ttfamily gr-qc/9905103}}].

\bibitem{Liu:2012xn}
H.-S.~Liu and H.~Lu, \emph{{Charged Rotating AdS Black Hole and Its
  Thermodynamics in Conformal Gravity}},
  \href{https://doi.org/10.1007/JHEP02(2013)139}{\emph{JHEP} {\bfseries 02}
  (2013) 139} [\href{https://arxiv.org/abs/1212.6264}{{\ttfamily 1212.6264}}].

\bibitem{Corral:2025npd}
C.~Corral, B.~Diez and E.~Papantonopoulos, \emph{{Nonconformally Ricci-flat
  instantons in conformal gravity with and without nonlinear matter fields}},
  \href{https://doi.org/10.1103/xgqn-z6f2}{\emph{Phys. Rev. D} {\bfseries 113}
  (2026) 064015} [\href{https://arxiv.org/abs/2510.11854}{{\ttfamily
  2510.11854}}].

\bibitem{Lee:1990nz}
J.~Lee and R.M.~Wald, \emph{{Local symmetries and constraints}},
  \href{https://doi.org/10.1063/1.528801}{\emph{J. Math. Phys.} {\bfseries 31}
  (1990) 725}.

\bibitem{Wald:1993nt}
R.M.~Wald, \emph{{Black hole entropy is the Noether charge}},
  \href{https://doi.org/10.1103/PhysRevD.48.R3427}{\emph{Phys. Rev. D}
  {\bfseries 48} (1993) R3427}
  [\href{https://arxiv.org/abs/gr-qc/9307038}{{\ttfamily gr-qc/9307038}}].

\bibitem{Iyer:1994ys}
V.~Iyer and R.M.~Wald, \emph{{Some properties of Noether charge and a proposal
  for dynamical black hole entropy}},
  \href{https://doi.org/10.1103/PhysRevD.50.846}{\emph{Phys. Rev. D} {\bfseries
  50} (1994) 846} [\href{https://arxiv.org/abs/gr-qc/9403028}{{\ttfamily
  gr-qc/9403028}}].

\bibitem{Wald:1999wa}
R.M.~Wald and A.~Zoupas, \emph{{A General definition of 'conserved quantities'
  in general relativity and other theories of gravity}},
  \href{https://doi.org/10.1103/PhysRevD.61.084027}{\emph{Phys. Rev. D}
  {\bfseries 61} (2000) 084027}
  [\href{https://arxiv.org/abs/gr-qc/9911095}{{\ttfamily gr-qc/9911095}}].

\bibitem{Mannheim:1988dj}
P.D.~Mannheim and D.~Kazanas, \emph{{Exact Vacuum Solution to Conformal Weyl
  Gravity and Galactic Rotation Curves}},
  \href{https://doi.org/10.1086/167623}{\emph{Astrophys. J.} {\bfseries 342}
  (1989) 635}.

\bibitem{Edery:1997hu}
A.~Edery and M.B.~Paranjape, \emph{{Classical tests for Weyl gravity:
  Deflection of light and radar echo delay}},
  \href{https://doi.org/10.1103/PhysRevD.58.024011}{\emph{Phys. Rev. D}
  {\bfseries 58} (1998) 024011}
  [\href{https://arxiv.org/abs/astro-ph/9708233}{{\ttfamily
  astro-ph/9708233}}].

\bibitem{Yoon:2013rxa}
Y.~Yoon, \emph{{Problems with Mannheim{\textquoteright}s conformal gravity
  program}}, \href{https://doi.org/10.1103/PhysRevD.88.027504}{\emph{Phys. Rev.
  D} {\bfseries 88} (2013) 027504}
  [\href{https://arxiv.org/abs/1305.0163}{{\ttfamily 1305.0163}}].

\bibitem{Mannheim:2015gba}
P.D.~Mannheim, \emph{{Comment on {\textquotedblleft}Problems with
  Mannheim{\textquoteright}s conformal gravity program{\textquotedblright}}},
  \href{https://doi.org/10.1103/PhysRevD.93.068501}{\emph{Phys. Rev. D}
  {\bfseries 93} (2016) 068501}
  [\href{https://arxiv.org/abs/1506.02479}{{\ttfamily 1506.02479}}].

\bibitem{Hobson:2021vzg}
M.~Hobson and A.~Lasenby, \emph{{Conformal gravity does not predict flat galaxy
  rotation curves}},
  \href{https://doi.org/10.1103/PhysRevD.104.064014}{\emph{Phys. Rev. D}
  {\bfseries 104} (2021) 064014}
  [\href{https://arxiv.org/abs/2103.13451}{{\ttfamily 2103.13451}}].

\bibitem{Mannheim:2021fql}
P.D.~Mannheim, \emph{{Critique of the use of geodesics in astrophysics and
  cosmology}}, \href{https://doi.org/10.1088/1361-6382/ac8140}{\emph{Class.
  Quant. Grav.} {\bfseries 39} (2022) 245001}
  [\href{https://arxiv.org/abs/2105.08556}{{\ttfamily 2105.08556}}].

\bibitem{Mannheim:2021osa}
P.D.~Mannheim, \emph{{Structure of conformal gravity in the presence of a scale
  breaking scalar field}},
  \href{https://doi.org/10.1007/s10714-022-02982-9}{\emph{Gen. Rel. Grav.}
  {\bfseries 54} (2022) 99} [\href{https://arxiv.org/abs/2105.14679}{{\ttfamily
  2105.14679}}].

\bibitem{Strominger:1984zy}
A.~Strominger, G.T.~Horowitz and M.J.~Perry, \emph{{Instantons in Conformal
  Gravity}}, \href{https://doi.org/10.1016/0550-3213(84)90340-7}{\emph{Nucl.
  Phys. B} {\bfseries 238} (1984) 653}.

\bibitem{Deser:1983wb}
S.~Deser and R.I.~Nepomechie, \emph{{Electric - Magnetic Duality of Conformal
  Gravitation}},
  \href{https://doi.org/10.1016/0375-9601(83)90654-0}{\emph{Phys. Lett. A}
  {\bfseries 97} (1983) 329}.

\bibitem{Corral:2021xsu}
C.~Corral, G.~Giribet and R.~Olea, \emph{{Self-dual gravitational instantons in
  conformal gravity: Conserved charges and thermodynamics}},
  \href{https://doi.org/10.1103/PhysRevD.104.064026}{\emph{Phys. Rev. D}
  {\bfseries 104} (2021) 064026}
  [\href{https://arxiv.org/abs/2105.10574}{{\ttfamily 2105.10574}}].

\bibitem{Caldarelli:2011idw}
M.M.~Caldarelli, R.G.~Leigh, A.C.~Petkou, P.M.~Petropoulos, V.~Pozzoli and
  K.~Siampos, \emph{{Vorticity in holographic fluids}},
  \href{https://doi.org/10.22323/1.155.0076}{\emph{PoS} {\bfseries CORFU2011}
  (2011) 076} [\href{https://arxiv.org/abs/1206.4351}{{\ttfamily 1206.4351}}].

\bibitem{Leigh:2011au}
R.G.~Leigh, A.C.~Petkou and P.M.~Petropoulos, \emph{{Holographic
  Three-Dimensional Fluids with Nontrivial Vorticity}},
  \href{https://doi.org/10.1103/PhysRevD.85.086010}{\emph{Phys. Rev. D}
  {\bfseries 85} (2012) 086010}
  [\href{https://arxiv.org/abs/1108.1393}{{\ttfamily 1108.1393}}].

\bibitem{Leigh:2012jv}
R.G.~Leigh, A.C.~Petkou and P.M.~Petropoulos, \emph{{Holographic Fluids with
  Vorticity and Analogue Gravity}},
  \href{https://doi.org/10.1007/JHEP11(2012)121}{\emph{JHEP} {\bfseries 11}
  (2012) 121} [\href{https://arxiv.org/abs/1205.6140}{{\ttfamily 1205.6140}}].

\bibitem{Mukhopadhyay:2013gja}
A.~Mukhopadhyay, A.C.~Petkou, P.M.~Petropoulos, V.~Pozzoli and K.~Siampos,
  \emph{{Holographic perfect fluidity, Cotton energy-momentum duality and
  transport properties}},
  \href{https://doi.org/10.1007/JHEP04(2014)136}{\emph{JHEP} {\bfseries 04}
  (2014) 136} [\href{https://arxiv.org/abs/1309.2310}{{\ttfamily 1309.2310}}].

\bibitem{Kalamakis:2020aaj}
G.~Kalamakis, R.G.~Leigh and A.C.~Petkou, \emph{{Aspects of holography of
  Taub-NUT- AdS$_4$ spacetimes}},
  \href{https://doi.org/10.1103/PhysRevD.103.126012}{\emph{Phys. Rev. D}
  {\bfseries 103} (2021) 126012}
  [\href{https://arxiv.org/abs/2009.08022}{{\ttfamily 2009.08022}}].

\bibitem{Bobev:2016sap}
N.~Bobev, T.~Hertog and Y.~Vreys, \emph{{The NUTs and Bolts of Squashed
  Holography}}, \href{https://doi.org/10.1007/JHEP11(2016)140}{\emph{JHEP}
  {\bfseries 11} (2016) 140}
  [\href{https://arxiv.org/abs/1610.01497}{{\ttfamily 1610.01497}}].

\bibitem{Bobev:2017asb}
N.~Bobev, P.~Bueno and Y.~Vreys, \emph{{Comments on Squashed-sphere Partition
  Functions}}, \href{https://doi.org/10.1007/JHEP07(2017)093}{\emph{JHEP}
  {\bfseries 07} (2017) 093}
  [\href{https://arxiv.org/abs/1705.00292}{{\ttfamily 1705.00292}}].

\bibitem{Bueno:2018yzo}
P.~Bueno, P.A.~Cano, R.A.~Hennigar and R.B.~Mann, \emph{{Universality of
  Squashed-Sphere Partition Functions}},
  \href{https://doi.org/10.1103/PhysRevLett.122.071602}{\emph{Phys. Rev. Lett.}
  {\bfseries 122} (2019) 071602}
  [\href{https://arxiv.org/abs/1808.02052}{{\ttfamily 1808.02052}}].

\bibitem{Bueno:2020odt}
P.~Bueno, P.A.~Cano, R.A.~Hennigar, V.A.~Penas and A.~Ruip{\'e}rez,
  \emph{{Partition functions on slightly squashed spheres and flux
  parameters}}, \href{https://doi.org/10.1007/JHEP04(2020)123}{\emph{JHEP}
  {\bfseries 04} (2020) 123}
  [\href{https://arxiv.org/abs/2001.10020}{{\ttfamily 2001.10020}}].

\bibitem{Bobev:2023lkx}
N.~Bobev, J.~Hong and V.~Reys, \emph{{Large N partition functions of 3d
  holographic SCFTs}},
  \href{https://doi.org/10.1007/JHEP08(2023)119}{\emph{JHEP} {\bfseries 08}
  (2023) 119} [\href{https://arxiv.org/abs/2304.01734}{{\ttfamily
  2304.01734}}].

\bibitem{Parmentier:2026aqh}
K.~Parmentier and N.~Bobev, \emph{{CFTs on squashed spheres and the thermal
  effective action}},
  \href{https://doi.org/10.1007/JHEP10(2026)079}{\emph{JHEP} {\bfseries 10}
  (2026) 079} [\href{https://arxiv.org/abs/2606.30716}{{\ttfamily
  2606.30716}}].

\bibitem{Hawking:1998ct}
S.W.~Hawking, C.J.~Hunter and D.N.~Page, \emph{{Nut charge, anti-de Sitter
  space and entropy}},
  \href{https://doi.org/10.1103/PhysRevD.59.044033}{\emph{Phys. Rev. D}
  {\bfseries 59} (1999) 044033}
  [\href{https://arxiv.org/abs/hep-th/9809035}{{\ttfamily hep-th/9809035}}].

\bibitem{Dowker:1998pi}
J.S.~Dowker, \emph{{Effective actions on the squashed three sphere}},
  \href{https://doi.org/10.1088/0264-9381/16/6/323}{\emph{Class. Quant. Grav.}
  {\bfseries 16} (1999) 1937}
  [\href{https://arxiv.org/abs/hep-th/9812202}{{\ttfamily hep-th/9812202}}].

\bibitem{Emparan:1999pm}
R.~Emparan, C.V.~Johnson and R.C.~Myers, \emph{{Surface terms as counterterms
  in the AdS / CFT correspondence}},
  \href{https://doi.org/10.1103/PhysRevD.60.104001}{\emph{Phys. Rev. D}
  {\bfseries 60} (1999) 104001}
  [\href{https://arxiv.org/abs/hep-th/9903238}{{\ttfamily hep-th/9903238}}].

\bibitem{Chamblin:1998pz}
A.~Chamblin, R.~Emparan, C.V.~Johnson and R.C.~Myers, \emph{{Large N phases,
  gravitational instantons and the nuts and bolts of AdS holography}},
  \href{https://doi.org/10.1103/PhysRevD.59.064010}{\emph{Phys. Rev. D}
  {\bfseries 59} (1999) 064010}
  [\href{https://arxiv.org/abs/hep-th/9808177}{{\ttfamily hep-th/9808177}}].

\bibitem{Birmingham:2001vd}
D.~Birmingham and S.~Mokhtari, \emph{{The Cardy-Verlinde formula and
  Taub-Bolt-AdS space-times}},
  \href{https://doi.org/10.1016/S0370-2693(01)00527-5}{\emph{Phys. Lett. B}
  {\bfseries 508} (2001) 365}
  [\href{https://arxiv.org/abs/hep-th/0103108}{{\ttfamily hep-th/0103108}}].

\bibitem{Clarkson:2002uj}
R.~Clarkson, L.~Fatibene and R.B.~Mann, \emph{{Thermodynamics of
  (d+1)-dimensional NUT charged AdS space-times}},
  \href{https://doi.org/10.1016/S0550-3213(02)01143-4}{\emph{Nucl. Phys. B}
  {\bfseries 652} (2003) 348}
  [\href{https://arxiv.org/abs/hep-th/0210280}{{\ttfamily hep-th/0210280}}].

\bibitem{Martelli:2011fw}
D.~Martelli and J.~Sparks, \emph{{The gravity dual of supersymmetric gauge
  theories on a biaxially squashed three-sphere}},
  \href{https://doi.org/10.1016/j.nuclphysb.2012.08.015}{\emph{Nucl. Phys. B}
  {\bfseries 866} (2013) 72} [\href{https://arxiv.org/abs/1111.6930}{{\ttfamily
  1111.6930}}].

\bibitem{Martelli:2012sz}
D.~Martelli, A.~Passias and J.~Sparks, \emph{{The supersymmetric NUTs and bolts
  of holography}},
  \href{https://doi.org/10.1016/j.nuclphysb.2013.04.026}{\emph{Nucl. Phys. B}
  {\bfseries 876} (2013) 810}
  [\href{https://arxiv.org/abs/1212.4618}{{\ttfamily 1212.4618}}].

\bibitem{Johnson:2014pwa}
C.V.~Johnson, \emph{{The Extended Thermodynamic Phase Structure of Taub-NUT and
  Taub-Bolt}},
  \href{https://doi.org/10.1088/0264-9381/31/22/225005}{\emph{Class. Quant.
  Grav.} {\bfseries 31} (2014) 225005}
  [\href{https://arxiv.org/abs/1406.4533}{{\ttfamily 1406.4533}}].

\bibitem{Johnson:2014xza}
C.V.~Johnson, \emph{{Thermodynamic Volumes for AdS-Taub-NUT and
  AdS-Taub-Bolt}},
  \href{https://doi.org/10.1088/0264-9381/31/23/235003}{\emph{Class. Quant.
  Grav.} {\bfseries 31} (2014) 235003}
  [\href{https://arxiv.org/abs/1405.5941}{{\ttfamily 1405.5941}}].

\bibitem{Bordo:2019slw}
A.B.~Bordo, F.~Gray and D.~Kubiz{\v{n}}{\'a}k, \emph{{Thermodynamics and Phase
  Transitions of NUTty Dyons}},
  \href{https://doi.org/10.1007/JHEP07(2019)119}{\emph{JHEP} {\bfseries 07}
  (2019) 119} [\href{https://arxiv.org/abs/1904.00030}{{\ttfamily
  1904.00030}}].

\bibitem{Diez:2026qjt}
B.~Diez, \emph{{Thermodynamics of Lorentzian Taub-NUT spacetimes in
  Einstein-Gauss-Bonnet AdS gravity}},
  \href{https://arxiv.org/abs/2609.13036}{{\ttfamily 2609.13036}}.

\bibitem{Hartnoll:2005yc}
S.A.~Hartnoll and S.P.~Kumar, \emph{{The $O(N)$ model on a squashed $S^3$ and
  the Klebanov-Polyakov correspondence}},
  \href{https://doi.org/10.1088/1126-6708/2005/06/012}{\emph{JHEP} {\bfseries
  06} (2005) 012} [\href{https://arxiv.org/abs/hep-th/0503238}{{\ttfamily
  hep-th/0503238}}].

\bibitem{Sezgin:2002rt}
E.~Sezgin and P.~Sundell, \emph{{Massless higher spins and holography}},
  \href{https://doi.org/10.1016/S0550-3213(02)00739-3}{\emph{Nucl. Phys. B}
  {\bfseries 644} (2002) 303}
  [\href{https://arxiv.org/abs/hep-th/0205131}{{\ttfamily hep-th/0205131}}].

\bibitem{Klebanov:2002ja}
I.R.~Klebanov and A.M.~Polyakov, \emph{{AdS dual of the critical O(N) vector
  model}}, \href{https://doi.org/10.1016/S0370-2693(02)02980-5}{\emph{Phys.
  Lett. B} {\bfseries 550} (2002) 213}
  [\href{https://arxiv.org/abs/hep-th/0210114}{{\ttfamily hep-th/0210114}}].

\bibitem{Bueno:2018uoy}
P.~Bueno, P.A.~Cano, R.A.~Hennigar and R.B.~Mann, \emph{{NUTs and bolts beyond
  Lovelock}}, \href{https://doi.org/10.1007/JHEP10(2018)095}{\emph{JHEP}
  {\bfseries 10} (2018) 095}
  [\href{https://arxiv.org/abs/1808.01671}{{\ttfamily 1808.01671}}].

\bibitem{Nambu:1961tp}
Y.~Nambu and G.~Jona-Lasinio, \emph{{Dynamical Model of Elementary Particles
  Based on an Analogy with Superconductivity. 1.}},
  \href{https://doi.org/10.1103/PhysRev.122.345}{\emph{Phys. Rev.} {\bfseries
  122} (1961) 345}.

\bibitem{Ashtekar:1984zz}
A.~Ashtekar and A.~Magnon, \emph{{Asymptotically anti-de Sitter space-times}},
  \href{https://doi.org/10.1088/0264-9381/1/4/002}{\emph{Class. Quant. Grav.}
  {\bfseries 1} (1984) L39}.

\bibitem{Ashtekar:1999jx}
A.~Ashtekar and S.~Das, \emph{{Asymptotically Anti-de Sitter space-times:
  Conserved quantities}},
  \href{https://doi.org/10.1088/0264-9381/17/2/101}{\emph{Class. Quant. Grav.}
  {\bfseries 17} (2000) L17}
  [\href{https://arxiv.org/abs/hep-th/9911230}{{\ttfamily hep-th/9911230}}].

\bibitem{Kouniatalis:2025pxs}
G.~Kouniatalis, P.A.~Gonz{\'a}lez, E.~Papantonopoulos and Y.~V{\'a}squez,
  \emph{{Near-horizon geodesic instabilities and anomalous decay of quasinormal
  modes in Weyl black holes}},
  \href{https://doi.org/10.1103/ftvr-pclv}{\emph{Phys. Rev. D} {\bfseries 112}
  (2025) 124061} [\href{https://arxiv.org/abs/2509.05164}{{\ttfamily
  2509.05164}}].

\bibitem{Olea:2005gb}
R.~Olea, \emph{{Mass, angular momentum and thermodynamics in four-dimensional
  Kerr-AdS black holes}},
  \href{https://doi.org/10.1088/1126-6708/2005/06/023}{\emph{JHEP} {\bfseries
  06} (2005) 023} [\href{https://arxiv.org/abs/hep-th/0504233}{{\ttfamily
  hep-th/0504233}}].

\bibitem{deHaro:2000vlm}
S.~de~Haro, S.N.~Solodukhin and K.~Skenderis, \emph{{Holographic reconstruction
  of space-time and renormalization in the AdS / CFT correspondence}},
  \href{https://doi.org/10.1007/s002200100381}{\emph{Commun. Math. Phys.}
  {\bfseries 217} (2001) 595}
  [\href{https://arxiv.org/abs/hep-th/0002230}{{\ttfamily hep-th/0002230}}].

\bibitem{Anastasiou:2020zwc}
G.~Anastasiou, O.~Miskovic, R.~Olea and I.~Papadimitriou, \emph{{Counterterms,
  Kounterterms, and the variational problem in AdS gravity}},
  \href{https://doi.org/10.1007/JHEP08(2020)061}{\emph{JHEP} {\bfseries 08}
  (2020) 061} [\href{https://arxiv.org/abs/2003.06425}{{\ttfamily
  2003.06425}}].

\bibitem{Taub:1950ez}
A.H.~Taub, \emph{{Empty space-times admitting a three parameter group of
  motions}}, \href{https://doi.org/10.2307/1969567}{\emph{Annals Math.}
  {\bfseries 53} (1951) 472}.

\bibitem{Newman:1963yy}
E.~Newman, L.~Tamburino and T.~Unti, \emph{{Empty space generalization of the
  Schwarzschild metric}}, \href{https://doi.org/10.1063/1.1704018}{\emph{J.
  Math. Phys.} {\bfseries 4} (1963) 915}.

\bibitem{Page:1985bq}
D.N.~Page and C.N.~Pope, \emph{{Inhomogeneous Einstein metrics on complex line
  bundles}}, \href{https://doi.org/10.1088/0264-9381/4/2/005}{\emph{Class.
  Quant. Grav.} {\bfseries 4} (1987) 213}.

\bibitem{Misner:1963fr}
C.W.~Misner, \emph{{The Flatter regions of Newman, Unti and Tamburino's
  generalized Schwarzschild space}},
  \href{https://doi.org/10.1063/1.1704019}{\emph{J. Math. Phys.} {\bfseries 4}
  (1963) 924}.

\bibitem{Atiyah:1975jf}
M.F.~Atiyah, V.K.~Patodi and I.M.~Singer, \emph{{Spectral asymmetry and
  Riemannian Geometry 1}},
  \href{https://doi.org/10.1017/S0305004100049410}{\emph{Math. Proc. Cambridge
  Phil. Soc.} {\bfseries 77} (1975) 43}.

\bibitem{Atiyah:1976jg}
M.F.~Atiyah, V.K.~Patodi and I.M.~Singer, \emph{{Spectral asymmetry and
  Riemannian geometry 2}},
  \href{https://doi.org/10.1017/S0305004100051872}{\emph{Math. Proc. Cambridge
  Phil. Soc.} {\bfseries 78} (1976) 405}.

\bibitem{Atiyah:1976qjr}
M.F.~Atiyah, V.K.~Patodi and I.M.~Singer, \emph{{Spectral asymmetry and
  Riemannian geometry. III}},
  \href{https://doi.org/10.1017/S0305004100052105}{\emph{Math. Proc. Cambridge
  Phil. Soc.} {\bfseries 79} (1976) 71}.

\bibitem{FeffermanGraham1985}
C.~Fefferman and C.R.~Graham, \emph{Conformal invariants},  in \emph{\'{E}lie
  Cartan et les math\'{e}matiques d'aujourd'hui, Lyon, 25--29 juin 1984},
  no.~S131 in Ast\'{e}risque, pp.~95--116, Soci\'{e}t\'{e} math\'{e}matique de
  France, 1985.

\bibitem{FeffermanGraham2012}
C.~Fefferman and C.R.~Graham, \emph{The Ambient Metric}, no.~178 in Annals of
  Mathematics Studies, Princeton University Press, Princeton, NJ (2012).

\bibitem{Balasubramanian:1999re}
V.~Balasubramanian and P.~Kraus, \emph{{A Stress tensor for Anti-de Sitter
  gravity}}, \href{https://doi.org/10.1007/s002200050764}{\emph{Commun. Math.
  Phys.} {\bfseries 208} (1999) 413}
  [\href{https://arxiv.org/abs/hep-th/9902121}{{\ttfamily hep-th/9902121}}].

\bibitem{Bianchi:2001kw}
M.~Bianchi, D.Z.~Freedman and K.~Skenderis, \emph{{Holographic
  renormalization}},
  \href{https://doi.org/10.1016/S0550-3213(02)00179-7}{\emph{Nucl. Phys. B}
  {\bfseries 631} (2002) 159}
  [\href{https://arxiv.org/abs/hep-th/0112119}{{\ttfamily hep-th/0112119}}].

\bibitem{Skenderis:2002wp}
K.~Skenderis, \emph{{Lecture notes on holographic renormalization}},
  \href{https://doi.org/10.1088/0264-9381/19/22/306}{\emph{Class. Quant. Grav.}
  {\bfseries 19} (2002) 5849}
  [\href{https://arxiv.org/abs/hep-th/0209067}{{\ttfamily hep-th/0209067}}].

\bibitem{Starobinsky:1982mr}
A.A.~Starobinsky, \emph{{Isotropization of arbitrary cosmological expansion
  given an effective cosmological constant}}, {\emph{JETP Lett.} {\bfseries 37}
  (1983) 66}.

\bibitem{Frodden:2021ces}
E.~Frodden and D.~Hidalgo, \emph{{The first law for the Kerr-NUT spacetime}},
  \href{https://doi.org/10.1016/j.physletb.2022.137264}{\emph{Phys. Lett. B}
  {\bfseries 832} (2022) 137264}
  [\href{https://arxiv.org/abs/2109.07715}{{\ttfamily 2109.07715}}].

\bibitem{Durka:2019ajz}
R.~Durka, \emph{{The first law of black hole thermodynamics for
  Taub{\textendash}NUT spacetime}},
  \href{https://doi.org/10.1142/S0218271822500213}{\emph{Int. J. Mod. Phys. D}
  {\bfseries 31} (2022) 2250021}
  [\href{https://arxiv.org/abs/1908.04238}{{\ttfamily 1908.04238}}].

\bibitem{Kalamakis:2025daq}
G.~Kalamakis and A.C.~Petkou, \emph{{Remarks on the quasinormal modes of
  Taub-NUT-AdS$_{4}$}},
  \href{https://doi.org/10.1088/1361-6382/ae1e54}{\emph{Class. Quant. Grav.}
  {\bfseries 42} (2025) 235015}
  [\href{https://arxiv.org/abs/2503.21001}{{\ttfamily 2503.21001}}].

\bibitem{Yonge:2006tn}
M.D.~Yonge, \emph{{AdS Taub-Nut space and the O(N) vector model on a squashed
  3-sphere}}, \href{https://doi.org/10.1088/1126-6708/2007/07/004}{\emph{JHEP}
  {\bfseries 07} (2007) 004}
  [\href{https://arxiv.org/abs/hep-th/0611154}{{\ttfamily hep-th/0611154}}].

\bibitem{Kastor:2009wy}
D.~Kastor, S.~Ray and J.~Traschen, \emph{{Enthalpy and the Mechanics of AdS
  Black Holes}},
  \href{https://doi.org/10.1088/0264-9381/26/19/195011}{\emph{Class. Quant.
  Grav.} {\bfseries 26} (2009) 195011}
  [\href{https://arxiv.org/abs/0904.2765}{{\ttfamily 0904.2765}}].

\bibitem{Kubiznak:2016qmn}
D.~Kubiznak, R.B.~Mann and M.~Teo, \emph{{Black hole chemistry: thermodynamics
  with Lambda}}, \href{https://doi.org/10.1088/1361-6382/aa5c69}{\emph{Class.
  Quant. Grav.} {\bfseries 34} (2017) 063001}
  [\href{https://arxiv.org/abs/1608.06147}{{\ttfamily 1608.06147}}].

\bibitem{Xu:2018vxf}
H.~Xu and M.-H.~Yung, \emph{{On the thermodynamic phase structure of conformal
  gravity}}, \href{https://doi.org/10.1016/j.physletb.2018.06.045}{\emph{Phys.
  Lett. B} {\bfseries 783} (2018) 36}
  [\href{https://arxiv.org/abs/1804.10446}{{\ttfamily 1804.10446}}].

\bibitem{Miskovic:2009bm}
O.~Miskovic and R.~Olea, \emph{{Topological regularization and self-duality in
  four-dimensional anti-de Sitter gravity}},
  \href{https://doi.org/10.1103/PhysRevD.79.124020}{\emph{Phys. Rev. D}
  {\bfseries 79} (2009) 124020}
  [\href{https://arxiv.org/abs/0902.2082}{{\ttfamily 0902.2082}}].

\bibitem{Irakleidou:2015exd}
M.~Irakleidou and I.~Lovrekovic, \emph{{Conformal gravity one-loop partition
  function}}, \href{https://doi.org/10.1103/PhysRevD.93.104043}{\emph{Phys.
  Rev. D} {\bfseries 93} (2016) 104043}
  [\href{https://arxiv.org/abs/1512.07130}{{\ttfamily 1512.07130}}].

\bibitem{Britto-Pacumio:1999dwj}
R.~Britto-Pacumio, A.~Strominger and A.~Volovich, \emph{{Holography for coset
  spaces}}, \href{https://doi.org/10.1088/1126-6708/1999/11/013}{\emph{JHEP}
  {\bfseries 11} (1999) 013}
  [\href{https://arxiv.org/abs/hep-th/9905211}{{\ttfamily hep-th/9905211}}].

\bibitem{Ghodsi:2014hua}
A.~Ghodsi, B.~Khavari and A.~Naseh, \emph{{Holographic Two-Point Functions in
  Conformal Gravity}},
  \href{https://doi.org/10.1007/JHEP01(2015)137}{\emph{JHEP} {\bfseries 01}
  (2015) 137} [\href{https://arxiv.org/abs/1411.3158}{{\ttfamily 1411.3158}}].

\bibitem{Godazgar:2019dkh}
H.~Godazgar, M.~Godazgar and C.N.~Pope, \emph{{Dual gravitational charges and
  soft theorems}}, \href{https://doi.org/10.1007/JHEP10(2019)123}{\emph{JHEP}
  {\bfseries 10} (2019) 123}
  [\href{https://arxiv.org/abs/1908.01164}{{\ttfamily 1908.01164}}].

\bibitem{Crawley:2023brz}
E.~Crawley, A.~Guevara, E.~Himwich and A.~Strominger, \emph{{Self-dual black
  holes in celestial holography}},
  \href{https://doi.org/10.1007/JHEP09(2023)109}{\emph{JHEP} {\bfseries 09}
  (2023) 109} [\href{https://arxiv.org/abs/2302.06661}{{\ttfamily
  2302.06661}}].

\bibitem{Doran:2026bng}
G.~Doran, R.~Monteiro and N.~Moynihan, \emph{{Memory effect from the scattering
  of Taub-NUT black holes}},
  \href{https://doi.org/10.1007/JHEP07(2026)188}{\emph{JHEP} {\bfseries 07}
  (2026) 188} [\href{https://arxiv.org/abs/2603.24365}{{\ttfamily
  2603.24365}}].

\bibitem{Heuveline:2026nxq}
S.~Heuveline, N.R.~Del~Pino and A.~Strominger, \emph{{The Conformal
  $w$-Algebra}},  \href{https://arxiv.org/abs/2609.13381}{{\ttfamily
  2609.13381}}.

\bibitem{Vasiliev:1990en}
M.A.~Vasiliev, \emph{{Consistent equation for interacting gauge fields of all
  spins in (3+1)-dimensions}},
  \href{https://doi.org/10.1016/0370-2693(90)91400-6}{\emph{Phys. Lett. B}
  {\bfseries 243} (1990) 378}.

\bibitem{Vasiliev:1992av}
M.A.~Vasiliev, \emph{{More on equations of motion for interacting massless
  fields of all spins in (3+1)-dimensions}},
  \href{https://doi.org/10.1016/0370-2693(92)91457-K}{\emph{Phys. Lett. B}
  {\bfseries 285} (1992) 225}.

\bibitem{Vasiliev:1999ba}
M.A.~Vasiliev, \emph{{Higher spin gauge theories: Star product and AdS space}},
   \href{https://arxiv.org/abs/hep-th/9910096}{{\ttfamily hep-th/9910096}}.

\bibitem{Vasiliev:2003ev}
M.A.~Vasiliev, \emph{{Nonlinear equations for symmetric massless higher spin
  fields in (A)dS(d)}},
  \href{https://doi.org/10.1016/S0370-2693(03)00872-4}{\emph{Phys. Lett. B}
  {\bfseries 567} (2003) 139}
  [\href{https://arxiv.org/abs/hep-th/0304049}{{\ttfamily hep-th/0304049}}].

\bibitem{Iazeolla:2007wt}
C.~Iazeolla, E.~Sezgin and P.~Sundell, \emph{{Real forms of complex higher spin
  field equations and new exact solutions}},
  \href{https://doi.org/10.1016/j.nuclphysb.2007.08.002}{\emph{Nucl. Phys. B}
  {\bfseries 791} (2008) 231}
  [\href{https://arxiv.org/abs/0706.2983}{{\ttfamily 0706.2983}}].

\bibitem{Iazeolla:2011cb}
C.~Iazeolla and P.~Sundell, \emph{{Families of exact solutions to Vasiliev's 4D
  equations with spherical, cylindrical and biaxial symmetry}},
  \href{https://doi.org/10.1007/JHEP12(2011)084}{\emph{JHEP} {\bfseries 12}
  (2011) 084} [\href{https://arxiv.org/abs/1107.1217}{{\ttfamily 1107.1217}}].

\bibitem{Skvortsov:2025ohi}
E.~Skvortsov and Y.~Yin, \emph{{Higher-spins on Taub-NUT and higher-spin
  Taub-NUT}}, \href{https://doi.org/10.1007/JHEP12(2025)099}{\emph{JHEP}
  {\bfseries 12} (2025) 099}
  [\href{https://arxiv.org/abs/2508.18804}{{\ttfamily 2508.18804}}].

\bibitem{Diaz:2024iuz}
F.~Diaz, C.~Iazeolla and P.~Sundell, \emph{{Fractional spins, unfolding, and
  holography. Part II. 4D higher spin gravity and 3D conformal dual}},
  \href{https://doi.org/10.1007/JHEP10(2024)066}{\emph{JHEP} {\bfseries 10}
  (2024) 066} [\href{https://arxiv.org/abs/2403.02301}{{\ttfamily
  2403.02301}}].

\bibitem{Diaz:2024kpr}
F.~Diaz, C.~Iazeolla and P.~Sundell, \emph{{Fractional spins, unfolding, and
  holography. Part I. Parent field equations for dual higher-spin gravity
  reductions}}, \href{https://doi.org/10.1007/JHEP09(2024)109}{\emph{JHEP}
  {\bfseries 09} (2024) 109}
  [\href{https://arxiv.org/abs/2403.02283}{{\ttfamily 2403.02283}}].

\bibitem{Vasiliev:2012vf}
M.A.~Vasiliev, \emph{{Holography, Unfolding and Higher-Spin Theory}},
  \href{https://doi.org/10.1088/1751-8113/46/21/214013}{\emph{J. Phys. A}
  {\bfseries 46} (2013) 214013}
  [\href{https://arxiv.org/abs/1203.5554}{{\ttfamily 1203.5554}}].

\bibitem{Maldacena:2011jn}
J.~Maldacena and A.~Zhiboedov, \emph{{Constraining Conformal Field Theories
  with A Higher Spin Symmetry}},
  \href{https://doi.org/10.1088/1751-8113/46/21/214011}{\emph{J. Phys. A}
  {\bfseries 46} (2013) 214011}
  [\href{https://arxiv.org/abs/1112.1016}{{\ttfamily 1112.1016}}].

\bibitem{Bonezzi:2016ttk}
R.~Bonezzi, N.~Boulanger, E.~Sezgin and P.~Sundell,
  \emph{{Frobenius{\textendash}Chern{\textendash}Simons gauge theory}},
  \href{https://doi.org/10.1088/1751-8121/50/5/055401}{\emph{J. Phys. A}
  {\bfseries 50} (2017) 055401}
  [\href{https://arxiv.org/abs/1607.00726}{{\ttfamily 1607.00726}}].

\bibitem{Boulanger:2015kfa}
N.~Boulanger, E.~Sezgin and P.~Sundell, \emph{{4D Higher Spin Gravity with
  Dynamical Two-Form as a Frobenius-Chern-Simons Gauge Theory}},
  \href{https://arxiv.org/abs/1505.04957}{{\ttfamily 1505.04957}}.

\bibitem{Parker}
J.R.~Parker, ``Notes on complex hyperbolic geometry.''
  \url{https://www.maths.dur.ac.uk/users/j.r.parker/img/NCHG.pdf}, 2003.

\end{thebibliography}\endgroup

\end{document}